\documentclass[lettersize,journal]{IEEEtran}

\usepackage{tikz,pgfplots}
\usepackage{mathdots}
\usepackage{yhmath}
\usepackage{cancel}
\usepackage{color}
\usepackage{siunitx}
\usepackage{array}
\usepackage{multirow}
\usepackage{amssymb}
\usepackage{gensymb}
\usepackage{tabularx}
\usepackage{extarrows}
\usepackage{booktabs}
\usetikzlibrary{fadings}
\usetikzlibrary{patterns}
\usetikzlibrary{shadows.blur}
\usetikzlibrary{shapes}
\usepackage{soul}

\usepackage{amsmath}
\usepackage{amsthm}
\usepackage{comment}
\usepackage{float}

\usepackage{flushend}
\usetikzlibrary{arrows}
\usepackage[hidelinks]{hyperref}  

\usepackage{environ}         
\usepackage{etoolbox}        
\usepackage{graphicx}        

\usepackage{caption}
\usepackage{subcaption}
\usepackage{float}

\newtheorem{remark}{Remark}

\begin{document}



\title{Learning-Enhanced Composite DNA Data Storage Under Sampling Randomness and IDS Errors}

\author{Busra~Tegin,~\IEEEmembership{Member,~IEEE,}
        and~Tolga M~Duman,~\IEEEmembership{Fellow,~IEEE}
\thanks{This work was funded by the European Union through the ERC Advanced
Grant 101054904: “TRANCIDS: Transmission over Channels with Insertions
and Deletions.” \\
A preliminary version of this work was presented at the 2026 IEEE International Conference on Communications (ICC) \cite{tegin2026robust}. }
\thanks{Busra Tegin is with  CentraleSupélec Rennes Campus, 35576 Cesson-Sevigne and IETR-UMR CNRS 6164, France
(e-mail: busra.tegin@centralesupelec.fr).}
\thanks{Tolga M. Duman is with the Department of Electrical and Electronics Engineering, Bilkent University, 06800 Ankara, Turkiye (e-mail:
duman@ee.bilkent.edu.tr).
}
}

\maketitle

\begin{abstract} 
DNA data storage offers a high-density, long-term alternative to conventional storage systems, addressing the exponential growth of digital data. Composite DNA extends this paradigm by leveraging mixtures of nucleotides to increase storage capacity beyond the four standard bases. In this work, composite DNA storage is modeled as a multinomial channel, and an analogy to digital modulation is established by representing composite letters on the three-dimensional probability simplex. To mitigate errors caused by sampling randomness, we derive transition probabilities for each constellation point which enables the computation of bit-wise log-likelihood ratios (LLRs) to employ practical channel codes for error correction. The framework is then extended to substitution and insertion-deletion-substitution (IDS) channels by proposing constellation update rules that account for these impairments. For the substitution channel, exact constellation updates enable exact LLR computation. In contrast, for the IDS channel, the large number of possible cases necessitates an approximate update rule, yielding approximate LLRs. To address this limitation, we further propose two learning-enhanced receiver architectures: (1) RhoNet, a learning-assisted model-based method that refines constellation points prior to analytical LLR computation, and (2) LLRNet, which directly estimates LLRs, representing a progressive transition from model-based analytical estimation to fully data-driven processing. Numerical results demonstrate reliable performance with existing low-density parity-check (LDPC) codes.
\end{abstract}


\section{Introduction}

The rapid growth of data generated by smartphones, smart homes, IoT devices, and other platforms is driving an exponential increase in digital information, projected to reach 384 zettabytes in 2028 \cite{Wright_IDC_2024_Datasphere}. As a result, the risk of a data storage crisis is rising, since traditional devices and methods cannot keep pace with this demand. To address this challenge, DNA data storage systems were introduced as early as the 1960s \cite{neiman1965molecular}. The core idea is to store binary data by encoding it into synthesized strands of DNA and later decoding it back. The practicality of DNA data storage was demonstrated in the early 2010s through large-scale 
proof-of-concept experiments
\cite{church2012next, goldman2013towards}, showing that it offers a high-density solution with remarkable durability \cite{erlich2017dna, grass2015robust}. In such systems, data is encoded using the four DNA bases $\{A,C,T,G\}$ to construct strands through DNA synthesis, which are then stored. To retrieve the stored data, a sequencing process is applied, where the strands are read and the original binary data is recovered.

Current synthesis and sequencing technologies are costly, and during the synthesis phase multiple 
copies of each DNA strand are inherently produced, resulting in significant redundancy. When reading 
(i.e., sequencing) these strands, one can exploit this redundancy by considering the multiple copies 
rather than treating them as duplicates to be discarded. To exploit this redundancy, composite DNA letters were introduced in \cite{anavy2019data}. In a composite letter, instead of using a single base from $\{A,C,T,G\}$, information is distributed across multiple strands as a mixture of nucleotides at specific positions, thereby extending the alphabet of standard DNA data storage beyond the four bases. A composite letter is characterized by a quartet of probabilities $[p_A, p_C, p_T, p_G]$, representing the expected frequency of each nucleotide across the strand copies. Since these probabilities sum to 1, only three values are sufficient to define a point in the 3-dimensional probability simplex.

Different aspects of composite DNA data storage systems have been studied in the literature. In \cite{9838471}, the channel impairments are modeled as limited-magnitude probability errors, and sphere packing and Gilbert-Varshamov bounds are derived, along with code constructions for this model. In \cite{10619202}, coding for composite DNA is explored under substitutions, strand losses, and deletions, and non-asymptotic upper bounds on the code size are obtained.  
In \cite{10619348}, the problem of selecting nucleotide mixtures that maximize decoding probability is addressed, and the expected number of reads required for reliable decoding is analyzed. For the multinomial channel, \cite{kobovich2023m} proposes an optimization algorithm to determine the capacity-achieving input distribution using the Blahut-Arimoto algorithm. In \cite{11107232}, a deep-learning based method for optimizing input distributions for composite DNA
under practical constraints such as noise and limited input support is presented. 
Extensions to combinatorial composite DNA, where shortmers replace base nucleotides, are studied in \cite{preuss2024efficient}, which proposes several encoding schemes and investigates their properties. A related work \cite{10619334} considers asymmetric errors in shortmer-based systems, where some shortmers may be missing from sequencing reads, and develops error-correcting codes for such scenarios.

Insertions, deletions, and substitutions are the most common error types observed in DNA data storage systems, and due to the synchronization loss caused by these errors, the design of the transmitter and receiver chains becomes highly challenging. 
For insertion-deletion-substitution channels, concatenated coding schemes are commonly employed to mitigate synchronization loss. In \cite{910582}, an outer low-density parity-check (LDPC) code providing error-correction capability is combined with an inner nonlinear watermark code for synchronization. A similar strategy is adopted in \cite{chen2003concatenated} for deletion channels, where the inner code consists of a combination of Varshamov-Tenengolts and marker codes. Likewise, \cite{5733455} utilizes an inner marker code for synchronization. 
Furthermore, the authors in \cite{5733455} introduce a symbol-level maximum a posteriori (MAP) detection algorithm that operates on groups of consecutive bits and demonstrate improvements in both the achievable rate and the error-rate performance.

In \cite{910582, chen2003concatenated}, similar concatenated coding schemes are proposed for channels with synchronization errors. As an alternative, data-driven solutions have also received increasing attention for complex channel models. More recently, for deletion and substitution channels, which are among the most common error types observed in DNA data storage systems, the authors of \cite{10622561,11358926} propose a deep-learning-based decoder designed for concatenated coding schemes, where bidirectional gated recurrent units (BiGRUs) are employed as log-likelihood ratio (LLR) estimators and outer code decoders are used to estimate the message bits. Furthermore, the authors propose a one-shot decoder for serial concatenation of inner marker code and outer convolutional code via a complete BiGRU based
architecture. Similarly, in \cite{10509546}, the decoding problem of marker codes for insertion and deletion channels is investigated under both known and unknown channel state information, such as insertion and deletion probabilities.

Beyond addressing the intractability arising from channel complexity, data-driven approaches can also reduce the computational burden of the receiver chain. Motivated by this advantage, several studies have employed machine-learning (ML)-based soft-output estimation methods. In \cite{9024433}, the authors propose a deep neural network approach to compute bit log-likelihood ratios (LLRs) of equalized symbols, achieving performance close to the optimal MAP solution with significantly reduced complexity. In \cite{9345504}, an ML-based digital radio receiver chain is proposed and trained in a supervised manner directly from frequency-domain antenna data, enabling not only soft demapping but also the integration of receiver blocks such as channel estimation and equalization for improved overall performance.

In this study, we consider a DNA data storage system with composite letters to exploit the redundancies inherent in the synthesis and sequencing processes. Following the approach in \cite{kobovich2023m}, we first define the multinomial channel for such systems and demonstrate a natural relationship between the constellation diagram of  a digital modulation scheme and composite DNA letters, represented on the three-dimensional probability simplex. Because of the randomness in DNA sampling, the observed frequencies may deviate from the original constellation points. To address this, we propose the use of practical error correcting codes, in particular LDPC codes, and derive the posterior probabilities of each constellation point for a given channel observation under
sampling randomness, thereby enabling the computation of
log-likelihood ratios.  
We further extend the proposed approach to commonly observed error models in DNA data storage systems, namely substitution and insertion–deletion–substitution (IDS) channels, and introduce a constellation update policy to account for these errors. For the substitution channel, since there is no synchronization loss among the nucleotides of the DNA strand, one can obtain an optimal constellation update policy, which results in exact LLR computations to be used by the channel decoder. However, when insertion and deletion errors are introduced into the system together with sampling randomness and substitution errors, synchronization loss occurs and the derivation of the optimal constellation update policy becomes intractable. Hence, we first propose an approximate and tractable constellation update method derived for small insertion and deletion probabilities, and these updated constellations are then used to obtain approximate LLRs for the considered channel model. The proposed approach is expected to provide solutions that are very close to the exact LLRs when IDS error probabilities are small, while it may significantly deviate from the exact LLRs for higher insertion and deletion probabilities. 
{\color{black}{To address this limitation and bridge the gap between analytical modeling and data-driven inference, we also propose two learning-enhanced soft information estimation approaches. Inspired by model-based deep learning \cite{shlezinger2023model}, these methods exploit partial domain knowledge while leveraging data-driven learning to handle complex synchronization errors. 
First, we introduce RhoNet, which employs a BiGRU architecture to estimate the updated constellation points, followed by analytical LLR computation. Building on this hybrid structure, we then propose LLRNet, which integrates the learning-based and analytical blocks into a unified framework and directly estimates the LLRs using a single BiGRU, without the explicit need for constellation updates. Similar bidirectional recurrent architectures have been successfully deployed for soft-information estimation and detection in storage and communication channels with memory, such as the multilayer BiGRU detector for magnetic recording channels in \cite{9279252} and the sliding bidirectional recurrent neural network for sequence detection in finite-memory channels in \cite{8454325}. 
Overall, the proposed framework provides a practical method that leverages the increased capacity of composite DNA while enabling robust error protection with existing LDPC codes.

Our main contributions can be summarized as follows:
\begin{itemize}
\item Unlike most existing studies on composite DNA storage, we consider the most commonly observed channel errors, namely substitution and insertion–deletion–substitution errors, together with sampling randomness, and propose a system model with improved robustness based on practical channel coding schemes, such as LDPC codes.
\item We provide an optimal constellation update policy for the substitution channel under the proposed composite DNA model with sampling randomness, resulting in exact LLR computations for channel decoding.
\item We extend the approach to the case of IDS errors, for which a suboptimal yet tractable analytical constellation-update policy is derived along with the corresponding approximate LLR calculations.
\item {\color{black}{To improve performance in the presence of IDS errors, particularly at higher error probabilities, we further propose two BiGRU-based receiver frameworks, namely RhoNet and LLRNet. These two methods represent a progressive transition from analytical solutions to data-driven processing and outperform the analytically derived suboptimal approach.}}
\item All the source code used to generate the numerical results has been provided in \cite{Busrategin2026}.
\end{itemize}

The paper is organized as follows. In Section~\ref{sec:system_model}, we introduce the system model and the underlying multinomial channel for DNA data storage systems with composite letters, incorporating channel-coding-based error protection. Section~\ref{sec:enc_dec} presents the encoding and decoding structures for the described system, along with the composite letter mapping and LLR computation for the multinomial composite DNA channel, as well as the introduction of the channel models. In Section~\ref{sec:analytical_llr_comp}, we extend the proposed LLR computation to the substitution channel with exact computations and to the IDS channel with approximate calculations. Section~\ref{sec:ml_llr} introduces our proposed BiGRU-based methods: RhoNet and LLRNet, for the IDS channel. The performance of composite DNA data storage for the proposed approaches with LDPC codes is presented in Section~\ref{sec:numerical_res}, and the paper is concluded in Section~\ref{sec:conc}.

\section{System Model} \label{sec:system_model}

We consider a DNA data storage system with composite letters, as illustrated in Fig.~\ref{fig:sm}. The input is a binary message sequence $\mathbf{m} = [m_1 \ m_2 \dots m_K]$, where each bit takes the value 0 or 1 with equal probability. This sequence is encoded using a channel code with rate $R = K/N$ to generate the codeword $\mathbf{c} = [c_1 \ c_2 \dots c_N]$, which is then mapped to composite DNA letters. A composite letter is defined as a mixture of the four standard DNA nucleotides $\{A, C, T, G\}$ in specified ratios at a given strand position. These composite letters exploit the redundancies inherent in DNA synthesis and sequencing, and are constructed by reading $n$ copies of a DNA strand at a specific index. The resulting channel is referred to as a multinomial channel, whose output is a quartet of empirical probabilities $[p_A, p_C, p_T, p_G]$ representing the observed frequency of each nucleotide in the $n$ copies.  

Each $L \in \mathbb{Z}^{+}$-bit segment of the codeword $\mathbf{c}$ is mapped to a composite letter according to a fixed constellation diagram. The input constellation is defined on the three-dimensional probability simplex:  
\begin{equation} \label{deltaL}
    \Delta_L = \left\{ \boldsymbol{\rho}_s \in \mathbb{R}_{+}^4 , \; \rho_{s,i} \in [0,1] \,\bigg|\, \sum_{i=1}^{4} \rho_{s,i} = 1 \right\},
\end{equation}
for $s \in \{1, \dots, S\}$ with $S = 2^L$. This set can be viewed as analogous to constellation points in standard digital modulation schemes. From the perspective of DNA synthesis, this corresponds to producing many strands, where at each position the nucleotide is chosen independently from $\{A, C, T, G\}$ with probabilities $\rho_{s,i}$, associated with the $s$-th constellation point. 
Hence, $\rho_{s,i}$ denotes the expected proportion of the $i$-th nucleotide synthesized at that specific position across the generated DNA strands (using the nucleotide order $A, C, T, G$).

\begin{figure}
    \centering
 \scalebox{0.53}{
\includegraphics[]{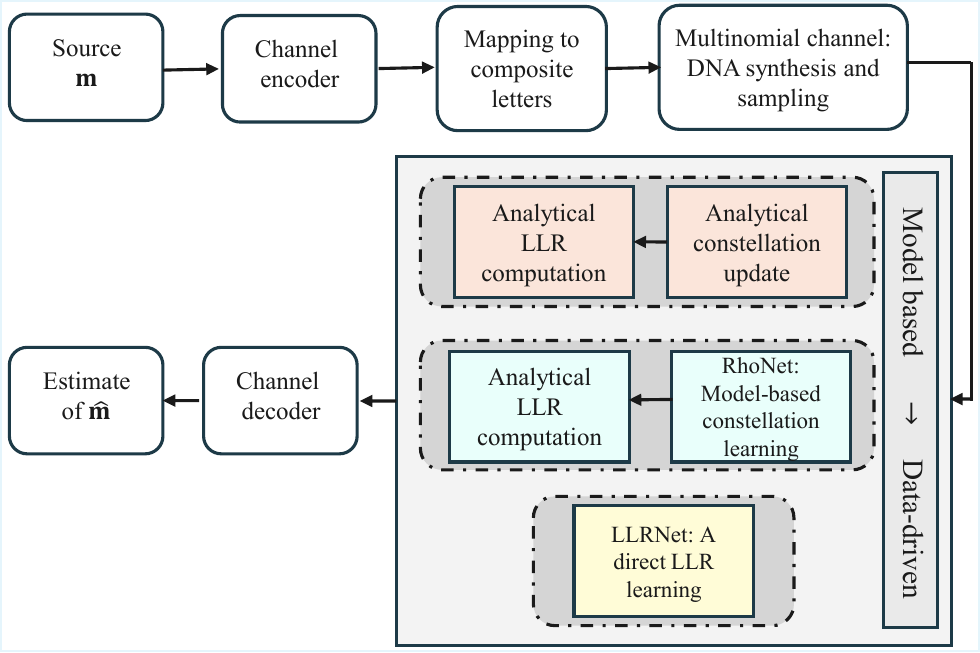}}
    \caption{System model for the composite DNA data storage system with channel coding based error protection.}
    \label{fig:sm}
\end{figure}

After the sampling process with $n$ reads (i.e., after reading $n$ DNA strands), the output of the multinomial channel can be represented by the matrix  
\begin{equation} \label{D1}
 \mathbf{D} = [\mathbf{d}_1 \ \mathbf{d}_2 \ \dots \ \mathbf{d}_E] \in \mathbb{R}_{+}^{3 \times E},   
\end{equation} 
where $\mathbf{d}_i \in \mathbb{R}_{+}^3$ for $i \in \{1, 2, \dots, E\}$, and $E = N/L$. For convenience, we assume that $N$ is an integer multiple of $L$.  

Consider the $i$-th observed symbol $\mathbf{d}_i = [d_{i,1} \ d_{i,2} \ d_{i,3}]^T$. This symbol can also be expressed as the ratio of 4-tuple $[k_A, k_C, k_T, k_G]$, where each entry corresponds to the number of observed nucleotides of type $A, C, T,$ and $G$, respectively. The relationships are given by  
\begin{equation} \label{nuc_num}
   \begin{aligned} 
      d_{i,1} &= p_A = \frac{k_A}{n}, \quad & d_{i,2} &= p_C = \frac{k_C}{n}, \\
      d_{i,3} &= p_T = \frac{k_T}{n}, \quad & d_{i,4} &= p_G = \frac{k_G}{n}.
   \end{aligned} 
\end{equation}
Since $n = k_A + k_C + k_T + k_G$, one of these quantities (e.g., $d_{i,4}$) can be omitted without loss of information, allowing us to work with only three values. The 4-tuple $[k_A, k_C, k_T, k_G]$ is determined by the synthesized sequences and is affected by sampling randomness as well as possible errors in DNA data storage systems, e.g., substitutions, deletions, and insertions. In the following, we focus only on sampling randomness and the resulting drawing error, and later extend 
the analysis to include substitutions, deletions, and insertions.

Due to the randomness in DNA sampling, the observed vectors $\mathbf{d}_i$ may deviate from the intended constellation points in $\Delta_L$, thereby introducing errors. Under this model, the 
channel effects can be determined using a multinomial distribution $\text{Multinomial}(n,\boldsymbol{\rho}_s)$. That is, given the input $\boldsymbol{\rho}_s \in \Delta_L$, the probability of observing the output $\mathbf{d}_i$ (without any impairments such as insertions, deletions, or substitutions) is
\begin{equation} \label{pdrho}
  P(\mathbf{d}_i \mid \boldsymbol{\rho}_s) 
  = \frac{n!}{\prod_{j=1}^{4} (n d_{i,j})!} 
    \prod_{j=1}^{4} \rho_{s,j}^{\, n d_{i,j}}.  
\end{equation}

Using these transition probabilities for all $S$ constellation points, we compute the LLRs for each composite letter, which are then used in the channel decoding process to estimate the original input bits.

\section{Encoding and Decoding Structures for Composite DNA and Channel Models} 
\label{sec:enc_dec}

In this section, we present the encoding structure, composite letter mapping, analytical LLR calculation, and decoding process of the proposed coding scheme for composite DNA data storage systems. The goal is to exploit the increased storage capacity of composite DNA letters while ensuring robustness against the randomness introduced by the sampling process.

\subsection{Encoding of the Binary Input Sequence}

Consider a binary input sequence of length $K$, denoted by $\mathbf{m}$. This sequence is encoded with a channel code of rate $R = K/N$ to generate the codeword $\mathbf{c}$.  
For example, when LDPC codes are employed, the generator matrix $\mathbf{G}$ has dimensions $K \times N$, and the  encoding process can be expressed as
\begin{equation}
    \mathbf{c} = \mathbf{m} \cdot \mathbf{G}.
\end{equation}
The resulting codeword $\mathbf{c}$ is then mapped to composite DNA letters from $\Delta_L$ as described in the next subsection.

\subsection{Mapping to DNA Composite Letters}

\begin{table}[]
\centering
\caption{An example of composite letter mapping with $L= 3$.} \label{tab:ex}
\begin{tabular}{|c|cccc|}
\hline
\multirow{2}{*}{$L$-bit input segments} & \multicolumn{4}{c|}{$\Delta_L$}                                                                                                          \\ \cline{2-5} 
                                        & \multicolumn{1}{c|}{$\rho_{s,1}$ (A)} & \multicolumn{1}{c|}{$\rho_{s,2}$ (C)} & \multicolumn{1}{c|}{$\rho_{s,3}$ (T)} & $\rho_{s,4}$ (G) \\ \hline
$s:1 \rightarrow \mathbf{u}_1=[000]$                                   & \multicolumn{1}{c|}{1}                & \multicolumn{1}{c|}{0}                & \multicolumn{1}{c|}{0}                & 0                \\ \hline
$s:2 \rightarrow \mathbf{u}_2=[001]$                                   & \multicolumn{1}{c|}{0}                & \multicolumn{1}{c|}{1}                & \multicolumn{1}{c|}{0}                & 0                \\ \hline
$s:3 \rightarrow \mathbf{u}_3=[010]$                                   & \multicolumn{1}{c|}{0}                & \multicolumn{1}{c|}{0}                & \multicolumn{1}{c|}{1}                & 0                \\ \hline
$s:4 \rightarrow \mathbf{u}_4=[011]$                                   & \multicolumn{1}{c|}{0}                & \multicolumn{1}{c|}{0}                & \multicolumn{1}{c|}{0}                & 1                \\ \hline
$s:5 \rightarrow \mathbf{u}_5=[100]$                                   & \multicolumn{1}{c|}{0.5}              & \multicolumn{1}{c|}{0.5}              & \multicolumn{1}{c|}{0}                & 0                \\ \hline
$s:6 \rightarrow \mathbf{u}_6=[101]$                                   & \multicolumn{1}{c|}{0}                & \multicolumn{1}{c|}{0}              & \multicolumn{1}{c|}{0.5}              & 0.5                \\ \hline
$s:7 \rightarrow \mathbf{u}_7=[110]$                                   & \multicolumn{1}{c|}{0.5}              & \multicolumn{1}{c|}{0}                & \multicolumn{1}{c|}{0.5}              & 0                \\ \hline
$s:8 \rightarrow \mathbf{u}_8=[111]$                                   & \multicolumn{1}{c|}{0}                & \multicolumn{1}{c|}{0.5}              & \multicolumn{1}{c|}{0}                & 0.5              \\ \hline
\end{tabular}
\end{table}

As described in Section~\ref{sec:system_model}, mapping the codeword $\mathbf{c}$ to composite letters is equivalent to mapping it onto three-dimensional constellation points. Specifically, each $L$-bit segment of $\mathbf{c}$ is associated with a constellation point $\boldsymbol{\rho}_s \in \Delta_L$, as defined in \eqref{deltaL}, where the corresponding input segment is denoted by $\mathbf{u}_s$, for all $s \in \{1, \dots, 2^L\}$.  

Table~\ref{tab:ex} illustrates a sample of mapping and constellation points for an exemplary system with $L = 3$. Although $\rho_{s,4}$ (corresponding to nucleotide G) is listed in the table, it is sufficient to specify only $[\rho_{s,1}, \rho_{s,2}, \rho_{s,3}]$ since their sum equals 1. Finally, note that the mapping of constellation points can be further optimized: the values of $\rho_{s,i}$ may be chosen based on the number of DNA copies $n$, as discussed in \cite{kobovich2023m}.

\begin{remark}
After this mapping, the information bits are represented by composite DNA letters corresponding to points in the probability simplex $\Delta_L$. From the perspective of DNA synthesis, many DNA strands are generated according to this mapping. For example, for the input bits `100' in Table~\ref{tab:ex}, the synthesis process produces strands such that half contain nucleotide `A' and the other half contain nucleotide `C'.
\end{remark}

\subsection{Exact LLR Computation Under Sampling Randomness}

After the mapping and subsequent DNA synthesis phase, the data is stored in DNA strands, which can later be read when retrieval is required. The reading process begins by obtaining $n$ samples of the DNA strands. As described in Section~\ref{sec:system_model}, the output of the multinomial channel is represented by the matrix \begin{equation}
    \mathbf{D} = [\mathbf{d}_1 \ \mathbf{d}_2 \ \dots \ \mathbf{d}_E],
\end{equation}
where the $i$-th observed symbol is $\mathbf{d}_i = [d_{i,1} \ d_{i,2} \ d_{i,3}]^T$ with the entries related to the number of observed nucleotides as defined in \eqref{nuc_num}.

To calculate the LLRs of the transmitted symbols given the channel observation, we first compute 
$P(\boldsymbol{\rho}_s \mid \mathbf{d}_i)$, i.e., the probability of each constellation point 
$\boldsymbol{\rho}_s$ given the observation vector $\mathbf{d}_i$, as
\begin{equation}
  P(\boldsymbol{\rho}_s \mid \mathbf{d}_i ) 
  = \frac{ P(\mathbf{d}_i \mid \boldsymbol{\rho}_s)\, P(\boldsymbol{\rho}_s)}{P(\mathbf{d}_i)}.
\end{equation}
For now, we only consider drawing errors (e.g., sampling randomness) and neglect other 
impairments such as substitutions, deletions, or insertions. Thus, one can directly use 
\eqref{pdrho}. Assuming that the prior probability $P(\boldsymbol{\rho}_s)$ is uniform across all constellation points, and since $P(\mathbf{d}_i)$ is identical for all composite symbols, we have
\begin{equation}
  P(\boldsymbol{\rho}_s \mid \mathbf{d}_i ) \propto P(\mathbf{d}_i \mid \boldsymbol{\rho}_s),
\end{equation}
which will then be used in the LLR calculation.

Let ${u}_{s,l}$ denote the $l$-th bit of $\mathbf{u}_{s}$ which is the $L$-bit segment defined by the mapping $\Delta_L$ for $\boldsymbol{\rho}_s$. 
The LLR corresponding to the $l$-th bit of the $i$-th symbol for $l \in \{1, \dots, L\}$ can be calculated as:
\begin{align} \label{eq_llr}
LLR_{i,l} &= \log \left( \frac{\sum\limits_{s: {u}_{s,l} = 0} P(\boldsymbol{\rho}_s \mid \mathbf{d}_i )}{\sum\limits_{s: {u}_{s,l} = 1} P(\boldsymbol{\rho}_s \mid \mathbf{d}_i )} \right) \nonumber \\
& = \log \left(  \frac{\sum\limits_{s: {u}_{s,l} = 0} P(\mathbf{d}_i \mid \boldsymbol{\rho}_s)}{\sum\limits_{s: {u}_{s,l} = 1} P(\mathbf{d}_i \mid \boldsymbol{\rho}_s)} \right) .
\end{align}
The resulting LLR vector 
\begin{align}
    \boldsymbol{\lambda} = [LLR_{1,1}, \dots,& LLR_{1,L}, \dots, \nonumber \\
    & \qquad LLR_{E,1}, \dots, LLR_{E,L}] \in \mathbb{R}^N,
\end{align}
is used as input to the channel decoder. For example, when LDPC codes are employed, any standard decoding algorithm for binary LDPC codes can be applied, such as the log-domain sum-product algorithm (log-SPA) or the min-sum algorithm \cite{ryan2009channel}.

\subsection{Channel Models Observed in DNA Data Storage Systems} \label{subsec:chs}
When composite letters are used in DNA data storage systems to increase their logical density (as described in Section \ref{sec:system_model}), the procedure is equivalent to mapping the codeword onto composite letters, which correspond to a fixed constellation diagram, and then synthesizing multiple DNA strands based on that mapping. To retrieve the stored information, one must sample a sufficient number of DNA strands, i.e., read $n$ strands, to estimate the fraction of each nucleotide, which is subsequently used to recover the message bits. 
As explained in the previous section, this process can be modeled using a multinomial channel when no additional errors are introduced; in this case, the sampling randomness is fully captured by that channel model.

However, DNA storage systems are also susceptible to additional error types, e.g.,  substitution, deletion, and insertion errors, which are summarized in the next section.

\subsubsection{Substitution Errors}
A commonly used mathematical framework for mutations in DNA data storage systems is the symmetric nucleotide substitution model. With this model, each nucleotide in a strand is assumed to be independently replaced by another nucleotide with equal probability, regardless of its position or neighboring nucleotides. However, we note that the practical synthesis and sequencing may tend to have skews in substitution rates, i.e., there may be variations between transitions and transversions. In any case, we consider a symmetric error distribution here for analytical tractability.

For this model, the channel transition probabilities are given by
\begin{align} \label{subs_ch}
       p_{i,j}= 
\begin{cases}
    1-p_s,& \text{if } i=j,\\[4pt]
    \tfrac{p_s}{3}, & \text{otherwise},
\end{cases}
\end{align}
where $i,j \in \{A,C,T,G\}$.

\subsubsection{Insertion-Deletion-Substitution Errors} \label{IDS_ch}
We consider IDS errors in each DNA strand under the following channel model: each nucleotide in a strand is either deleted with probability $p_d$, followed by at most a single random nucleotide insertion after its position with probability $p_i$ (each of the four nucleotides being equally likely with probability $p_i/4$), the nucleotide is transmitted incorrectly (substituted) with probability $ \bar{p}_s = (1 - p_d - p_i)p_s $ (each nucleotide is substituted with one of the other nucleotides, each with probability $\bar p_s/3$), or it is transmitted correctly with probability $(1 - p_d - p_i)(1 - p_s)$. This formulation represents a simplified insertion model by restricting insertions to a single event per position to guarantee analytical tractability, which is a widely adopted convention in recent DNA data storage channel models (see, e.g., \cite{MGC, 5733455, 9517821}).

Note that this general IDS model can also represent different combinations of the three error types, e.g., a deletion-only channel, an insertion-only channel, a substitution–deletion channel, or an insertion–deletion channel, by assigning nonzero probabilities to the error types of interest and setting the remaining ones to zero.

\begin{remark}
The insertion, deletion, and substitution error models introduced above represent a simplified abstraction of the mutations observed in physical DNA data storage systems. In practical scenarios, these errors are often context-dependent and position-dependent, affected by other factors, e.g., the local nucleotide composition (e.g., GC-content or homopolymer runs) and adjacent base configurations, leading to non-identical error distributions. Similarly, for the multinomial channel, the molecular sampling process may have biases rather than having a strict uniform distribution across strands. However, due to the lack of experimental datasets characterizing composite nucleotide mixtures, along with the need to preserve the mathematical tractability of the derived analytical LLR approach, we employ these standardized, uniform channel assumptions, which are widely accepted and well-established in the relevant DNA data storage literature.
\end{remark}

\section{Analytical LLR Computation} \label{sec:analytical_llr_comp}
When the only source of error is sampling randomness, one can compute the exact LLR of the composite letter using the formulation given in \eqref{eq_llr}. However, when additional error types are present, such as substitution, deletion, or insertion errors, the nucleotide observed at a given strand position may differ from the expected set. For instance, consider a position corresponding to input bits `100' represented by the composite letter $\boldsymbol{\rho_5} = [0.5, 0.5, 0, 0]$ (where $\rho_{5,1}=0.5$ for A, $\rho_{5,2}=0.5$ for C, and zero for T and G) as given in Table \ref{tab:ex}. Under sampling randomness alone, only reading nucleotides $A$ and $C$ would be expected. However, if a strand experiences a substitution error at that position, or an insertion/deletion error occurs before that index, unexpected nucleotides such as $T$ or $G$ may also be observed. More generally, channel-induced errors alter the expected observation frequencies of each nucleotide for the predefined constellation points. To account for this modified nucleotide distribution and preserve data recoverability, an updated composite letter representation, denoted by $\boldsymbol{\hat{\rho}}_s$, must be constructed.

In the next section, we derive the exact LLR formulation for the substitution channel, followed by approximations for the IDS channels. This extends our approach to obtain the LLRs by updating the corresponding constellation points accordingly.

\subsection{Substitution Error}

One of the most common mutations in DNA data storage is nucleotide substitution, where under the symmetric nucleotide substitution model, each nucleotide in a strand is independently replaced with a random nucleotide, regardless of its position or neighboring bases. As described in \eqref{subs_ch}, using the transition probabilities, each nucleotide is either transmitted correctly with probability $1 - p_s$ or substituted with one of the other nucleotides, each with probability $\tfrac{p_s}{3}$.

Due to the substitution error, when the strands are read, the probability $\rho_{s,i}$ of the $i$-th nucleotide will no longer accurately represent the fraction of observed nucleotides.  

We illustrate this with a simple example by considering two cases for nucleotide A:
\begin{enumerate}
    \item Suppose the synthesized base is A with probability $\rho_{s,1}$. It is transmitted correctly over the substitution channel with probability $1-p_s$, resulting in a contribution of $\rho_{s,1}(1-p_s)$ for nucleotide A.
    \item Suppose the synthesized base is not A with probability $1-\rho_{s,1}$, and it is substituted into A with probability $p_s/3$. Hence, the contribution for this case is $\tfrac{p_s}{3}(1-\rho_{s,1})$.
\end{enumerate}
Adding these two cases, we obtain $
\hat{\rho}_{s,1} = \rho_{s,1}(1-p_s) + \tfrac{p_s}{3}(1-\rho_{s,1})$. In general, for $j \in \{1,2,3,4\}$ one can write
\begin{equation} \label{upd_subs}
    \hat{\rho}_{s,j} = (1-p_s)\rho_{s,j} + \frac{p_s}{3}(1-\rho_{s,j}),
\end{equation}  
where $\boldsymbol{\hat{\rho}}_s = [\hat{\rho}_{s,1}, \ldots, \hat{\rho}_{s,4}]$, for $s \in \{1, \ldots, 2^L\}$. In vector form, this can be rewritten as
\begin{equation} \label{upd_subs2}
    \hat{\boldsymbol{\rho}}_{s} = \boldsymbol{\rho}_s \left(1- \tfrac{4}{3}p_s\right) + \tfrac{p_s}{3}[1 \ 1 \ 1 \ 1].
\end{equation}
Now consider $n$ independent reads. Each read is distributed according to $\hat{\boldsymbol{\rho}}_{s}$, and the observed counts $[k_A, k_C, k_T, k_G]$ follow the distribution $\text{Multinomial}\!\left(n,\hat{\boldsymbol{\rho}}_{s}\right)$.  
To address this, we update each constellation point as $\boldsymbol{\hat{\rho}}_s$, and the updated points $\boldsymbol{\hat{\rho}}_s$ are then used in place of $\boldsymbol{\rho}_s$ in \eqref{pdrho} and \eqref{eq_llr}.

Furthermore, when only sampling randomness is present, if a constellation point $\boldsymbol{\rho}_s$ has a zero entry, the probability of synthesizing or reading the corresponding nucleotide is also zero. For example, if $\boldsymbol{\rho}_s = [0.5, 0.5, 0, 0]$ (for A, C, T, and G, respectively), then in the absence of additional errors it is impossible to observe T or G.  
Thus, the observed letter $\mathbf{d}_i$ will contain zeros in these positions, e.g., $d_{i,3} = d_{i,4} = 0$. However, when substitution errors occur, nonzero values may appear in these positions within $\mathbf{d}_i$, as substitutions can introduce nucleotides not originally present in $\boldsymbol{\rho}_s$, or cause the overall observation probabilities to deviate significantly from the nominal constellation design. As a result, the standard LLR defined in \eqref{eq_llr} may no longer accurately reflect the true likelihood of the constellation points. Updating the constellation points as shown in \eqref{upd_subs2} resolves this discrepancy by explicitly accounting for these unexpected observations, thereby restoring the reliability of the LLR calculations.

\subsection{Insertion–Deletion-Substitution Errors}

We consider the definition of IDS channel as described in Section \ref{subsec:chs} with three possible error types: 1)
each nucleotide in a strand is either deleted with probability $p_d$, 2) followed by a random nucleotide insertion after its position with probability $p_i$, 3) the nucleotide is transmitted incorrectly (substituted) with probability $ \bar{p}_s = (1 - p_d - p_i)p_s $, or it is transmitted correctly with probability $(1 - p_d - p_i)(1 - p_s)$. We further define the equivalent substitution error model to the one introduced in Section~\ref{subsec:chs} in order to simplify the analytical calculations. A nucleotide is transmitted correctly with probability \(1-\hat{p}_s\), or substituted with one of \(\{A,C,T,G\}\) with equal probabilities \(\hat{p}_s/4\). Clearly, defining \(\hat{p}_s = 4p_s/3\) makes these two error models equivalent to each other. This transformation will be necessary in our analytical calculations, in which we use the core idea that a nucleotide will be replaced by one of the four possible nucleotides \(\{A,C,T,G\}\) if there is a position shift due to insertions or deletions, or a substitution at the given position.

Similar to the substitution channel, when the strand passes through the IDS channel, the transition probability in \eqref{pdrho} and the LLR in \eqref{eq_llr} may no longer reflect the true likelihood of the constellation points. Therefore, we update the constellation points accordingly.

We note that optimal computation is challenging due to the large number of possible combinations of insertions, deletions, and substitutions. Therefore, we restrict our attention to strands of length $E$, $E-1$, and $E+1$, i.e., strands with no insertions or deletions, or with an equal number of insertions and deletions, or with a difference of exactly one between the number of insertions and deletions, while discarding the remaining cases. Furthermore, the synchronization loss caused by channel errors may induce statistical dependencies and correlations among neighboring indices of the strand, which also affects the LLRs. In this work, these inter-symbol dependencies are ignored to preserve mathematical tractability. Hence, the analytical framework derived below provides a tractable but approximate approach for constellation updates, rather than an exact, optimal solution.

We first account for the contribution coming from strands of length $E$. Conditioned on observing such a strand, we compute the probability of having no shift/nucleotide substitution for position $i \in \{1, \ldots, E\}$ as
\begin{align}
   & p_{\text{ns},i,E} = P(\text{no shift/substitution for position } i \mid \mathcal{C}_E) \nonumber \\
    &= \frac{P(\text{no shift/subs. for position } i \text{ and event } \mathcal{C}_E)}{P(\mathcal{C}_E)},
\end{align}
where the event $\mathcal{C}_E$ denotes observing a strand of length $E$.
Since most DNA data storage systems employ short strands with small to moderate deletion and insertion probabilities \cite{heckel2019characterization, sabary2024reconstruction}, almost all strands will have at most one insertion or one deletion. Hence, we assume that if the strand length is $E$, then either there is no insertion/deletion, or there is exactly one insertion and one deletion. The probability of two or more insertions/deletions is assumed to be negligible.

Under this assumption, position $i$ experiences no shift or nucleotide change if (i) no insertions or deletions occur up to index $i-1$ and $i$, respectively, and no substitution occurs at index $i$, or (ii) exactly one insertion and one deletion occur up to index $i$, with no substitution at index $i$. Note that substitutions at positions other than index $i$ do not affect the nucleotide at position $i$. For the remaining calculations, instead of using the substitution channel model introduced in Section~\ref{subsec:chs}, we use the following equivalent transformation: a nucleotide is transmitted correctly with probability $1-\hat p_s$, or substituted with one of $\{A,C,T,G\}$ with equal probability $\hat p_s/4$. Clearly, $\hat{p}_s = 4p_s/3$ makes the two models equivalent. This yields
\begingroup
\allowdisplaybreaks
\begin{align} \label{prob1}
    &P(\text{no shift/nucleotide substitution for position } i \text{ and } \mathcal{C}_E) \nonumber \\
    &= \left(1-p_i-p_d\right)^{E-2} (1-\hat p_s) \Bigg( (1-p_i-p_d)^2 \nonumber \\
    &\qquad + p_i p_d \left(2\left( {E-i \choose 2} + {i-1 \choose 2} \right) + E-i\right)\Bigg),
\end{align}
\endgroup
where $\hat p_s = \frac{4p_s}{3}$. 
Here, the combinatorial terms explicitly account for the four mutually exclusive scenarios where position $i$ suffers no net shift or substitution:
\begin{itemize}
    \item \textbf{Scenario 1:} No insertions or deletions occur anywhere in the strand and no substitution in the position $i$, which corresponds to the term $\left(1-p_i-p_d\right)^{E} (1-\hat p_s)$.
    \item \textbf{Scenario 2:} Exactly one insertion and one deletion occur, both located after position $i$ with no substitution in the position $i$, yielding $2 p_i p_d \left(1-p_i-p_d\right)^{E-2} {E-i \choose 2} (1-\hat p_s)$.
    \item \textbf{Scenario 3:} Exactly one insertion and one deletion occur, both located before position $i$ with no substitution in the position $i$, yielding $2 p_i p_d \left(1-p_i-p_d\right)^{E-2} {i-1 \choose 2} (1-\hat p_s)$.
    \item \textbf{Scenario 4:} Exactly one insertion just after position $i$ and one complementary deletion occurs after position $i$ with no substitution in the position $i$, yielding $p_i p_d \left(1-p_i-p_d\right)^{E-2} (E-i) (1-\hat p_s)$.
\end{itemize}
We also have:
\begin{align}
    P(\mathcal{C}_E) = \left(1-p_i-p_d\right)^{E-2} \left( (1-p_i-p_d)^2 + 2p_i p_d {E \choose 2} \right),
\end{align}
where $\mathcal{C}_E$ represents the conditioning event that the received strand has length $E$. Note that nucleotide substitutions do not alter the length of the strand and are therefore omitted from this specific length probability. This expression accounts for two mutually exclusive cases:
\begin{itemize}
    \item \textbf{No insertions or deletions:} No indels occur anywhere across the $E$ positions of the strand, which corresponds to the term $\left(1-p_i-p_d\right)^{E}$.
    \item \textbf{Exactly one insertion and one deletion:} Exactly one insertion and one deletion occur within the strand (maintaining a net length change of zero), which yields the term $2 p_i p_d \left(1-p_i-p_d\right)^{E-2} {E \choose 2}$.
\end{itemize}
This gives
\begin{align} \label{noshift_E}
    p_{\text{ns},i,E} =
    (1-\hat p_s) \Bigg( &\frac{ (1-p_i-p_d)^2 }
         { (1-p_i-p_d)^2 + 2p_i p_d {E \choose 2} } \nonumber \\
         &+ \frac{ p_i p_d \left( 2 \left({E-i \choose 2} + {i-1 \choose 2} \right) +E-i\right) }
         { (1-p_i-p_d)^2 + 2p_i p_d {E \choose 2} } \Bigg).
\end{align}
Note that $p_{\text{ns},i,E}$ is defined for a single strand of length $E$.
If we observe $n_E \leq n$ strands of length $E$, then for position $i$, $t$ of the strands may experience a shift while the remaining $n_E - t$ do not. This occurs with probability
\begin{equation} \label{shift}
{n_E \choose t} p_{\text{ns},i,E}^{n_E - t} \left(1 - p_{\text{ns},i,E}\right)^t.
\end{equation}
Accordingly, we propose to update the constellation points as
\begin{align} \label{ID_upd2}
\boldsymbol{\hat{\rho}}_{s,E} &= \sum_{t=0}^{n_E} {n_E \choose t} p_{\text{ns},i,E}^{n_E - t} \left(1 - p_{\text{ns},i,E}\right)^t \nonumber \\
    &\qquad \cdot \Bigg[ 0.25\frac{t}{n_E}[1 \ 1 \ 1 \ 1] + \frac{n_E - t}{n_E} \boldsymbol{\rho}_s \Bigg].
\end{align}

We repeat the same derivation for strands of length $E-1$ and $E+1$, where we assume that a strand of length $E-1$ results from exactly one deletion and no insertion, and a strand of length $E+1$ results from exactly one insertion and no deletion. Events involving a larger number of insertions and deletions that lead to the same net length change, e.g., two deletions together with one insertion (or one deletion together with two insertions), are ignored, since they occur with much lower probability and for the tractability of the proposed approach.

For strands of length $E-1$, let $\mathcal{C}_{E-1}$ denote the event of observing a strand of length $E-1$, and let $n_{E-1}$ denote the number of such observed strands.
Conditioned on observing such a strand with length $E-1$, we compute the probability of having no shift/nucleotide substitution for position $i \in \{1, \ldots, E\}$ as
\begin{align}
   & p_{\text{ns},i,E-1} = P(\text{no shift/substitution for position } i \mid \mathcal{C}_{E-1}) \nonumber \\
    &= \frac{P(\text{no shift/subs. for position } i \text{ and event } \mathcal{C}_{E-1})}{P(\mathcal{C}_{E-1})}.
\end{align}
We have
\begin{align}
    &P(\text{no shift/substitution for position } i \text{ and } \mathcal{C}_{E-1}) \nonumber \\
    &= \left(1-p_i-p_d\right)^{E-1} (1-\hat p_s) \, p_d \left(E-i\right),
\end{align}
and
\begin{align}
    P(\mathcal{C}_{E-1}) = \left(1-p_i-p_d\right)^{E-1} p_d E,
\end{align}
which results in
\begin{align} \label{noshift_Em1}
    p_{\text{ns},i,E-1} = (1-\hat p_s) \, \frac{E-i}{E}.
\end{align}
Similar to the previous case, the corresponding constellation point is updated as
\begin{align} \label{ID_upd2E-1}
    \boldsymbol{\hat{\rho}}_{s,E-1} &= \sum_{t=0}^{n_{E-1}} {n_{E-1} \choose t} p_{\text{ns},i,E-1}^{\,n_{E-1} - t} \left(1 - p_{\text{ns},i,E-1}\right)^t \nonumber \\
    &\qquad \cdot \Bigg[ 0.25\frac{t}{n_{E-1}}[1 \ 1 \ 1 \ 1] + \frac{n_{E-1} - t}{n_{E-1}} \boldsymbol{\rho}_s \Bigg].
\end{align}

For strands of length $E+1$, let $\mathcal{C}_{E+1}$ denote the event of observing a strand of length $E+1$, and let $n_{E+1}$ denote the number of such observed strands. Conditioned on observing such a strand with length $E+1$, we define
\begin{align}
   & p_{\text{ns},i,E+1} = P(\text{no shift/substitution for position } i \mid \mathcal{C}_{E+1}) \nonumber \\
    &= \frac{P(\text{no shift/subs. for position } i \text{ and event } \mathcal{C}_{E+1})}{P(\mathcal{C}_{E+1})}.
\end{align}
Since we have
\begin{align}
    &P(\text{no shift/substitution for position } i \text{ and } \mathcal{C}_{E+1}) \nonumber \\
    &= \left(1-p_i-p_d\right)^{E-1} (1-\hat p_s) \, p_i \left(E-i+1\right),
\end{align}
and
\begin{align}
    P(\mathcal{C}_{E+1}) = \left(1-p_i-p_d\right)^{E-1} p_i E,
\end{align}
which gives
\begin{align} \label{noshift_Ep1}
    p_{\text{ns},i,E+1} = (1-\hat p_s) \, \frac{E-i+1}{E}.
\end{align}
Hence, the corresponding update rule for strands of length $E+1$ is
\begin{align} \label{ID_upd2-E+1}
\boldsymbol{\hat{\rho}}_{s,E+1} &= \sum_{t=0}^{n_{E+1}} {n_{E+1} \choose t} p_{\text{ns},i,E+1}^{\,n_{E+1} - t} \left(1 - p_{\text{ns},i,E+1}\right)^t \nonumber \\
    &\qquad \cdot \Bigg[ 0.25\frac{t}{n_{E+1}}[1 \ 1 \ 1 \ 1] + \frac{n_{E+1} - t}{n_{E+1}} \boldsymbol{\rho}_s \Bigg].
\end{align}

We note that, while $\boldsymbol{\hat{\rho}}_{s,E}$, $\boldsymbol{\hat{\rho}}_{s,E-1}$, and $\boldsymbol{\hat{\rho}}_{s,E+1}$ depend on the index $i$ of the corresponding position in the strand, for simplicity, we omit it in the notation.

Let $\mathcal{S}_{E}$, $\mathcal{S}_{E-1}$ and $\mathcal{S}_{E+1}$ denote the (disjoint) sets of
received strands whose lengths are $E$, $E-1$ and $E+1$, with cardinalities
$n_{E}$, $n_{E-1}$ and $n_{E+1}$, respectively, so that
$n_{E}+n_{E-1}+n_{E+1}\le n$, the remaining strands being discarded.
For each length class $\ell\in\{E,E-1,E+1\}$ and each position $i$, we form a
separate count vector
\begin{equation}
    \mathbf{d}_i^{(\ell)} = \big[k_A^{(\ell)},k_C^{(\ell)},k_T^{(\ell)},k_G^{(\ell)}\big]/n_{\ell},
\end{equation}
by counting only the nucleotides observed at index $i$ of the strands in
$\mathcal{S}_{\ell}$. Hence each received strand contributes to exactly one of the three
count vectors and no observation is used more than once.
The LLR in \eqref{eq_llr} are then evaluated three times,
once per length class, using $\hat{\boldsymbol{\rho}}_{s,\ell}$ in place of
$\boldsymbol{\rho}_{s}$, $\mathbf{d}_i^{(\ell)}$ in place of $\mathbf{d}_i$, and
$n_{\ell}$ in place of $n$. Conditioned on the transmitted composite letter, the three
subsets are independent, so the resulting soft outputs add in the log domain,
\begin{equation} \label{final_LLR}
    \text{LLR}_{i,l} = \text{LLR}_{i,l}(\hat{\boldsymbol{\rho}}_{s,E})
    + \text{LLR}_{i,l}(\hat{\boldsymbol{\rho}}_{s,E-1})
    + \text{LLR}_{i,l}(\hat{\boldsymbol{\rho}}_{s,E+1}),
\end{equation}
where $\text{LLR}_{i,l}(\hat{\boldsymbol{\rho}}_{s,\ell})$ is computed from
$\mathbf{d}_i^{(\ell)}$.

Furthermore, it should be noted that since the proposed approach discards the strands whose lengths are different from $E$, $E-1$, and $E+1$, and also ignores the dependencies between neighboring positions, there is an information loss, and the proposed constellation point update rule constitutes an approximation rather than an optimal solution for LLR calculation. Hence, the proposed method is expected to perform better in the low-error regime, which is a realistic assumption for practical DNA data storage systems \cite{heckel2019characterization}.

\begin{remark}
    In \eqref{ID_upd2}, \eqref{ID_upd2E-1}, and \eqref{ID_upd2-E+1}, we assume that when synchronization is lost, the shifted nucleotide is uniformly distributed over the alphabet, and thus we use $0.25[1 \ 1 \ 1 \ 1]$ since the underlying constellation design is symmetric. For asymmetric designs, this term can be updated according to the specific weights assigned to different nucleotides.
\end{remark}

\begin{remark}
This approach can be easily extended to a deletion-only channel, an insertion-only channel, an insertion–deletion channel, or any other combination of insertion, deletion, and substitution errors by setting the probabilities of the non-observed error types to zero in \eqref{prob1}-\eqref{final_LLR}.
\end{remark}

\begin{remark}
When only substitution errors are present in the channel, all sampled strands have the same length as the original strand, resulting in $n_E = n$ and $n_{E-1} = n_{E+1} = 0$. Furthermore, by setting $p_d = p_i = 0$ and defining
$p_{\text{ns},i,E} = 1-\hat p_s = 1- \frac{4p_s}{3}$, and using
\[
\sum_{t=0}^{n} \binom{n}{t} (1-\hat p_s)^{\,n-t}(\hat p_s)^t \, t \;=\; n\hat p_s,
\]
one can easily show that
\begin{align} \label{ID_upd2subs}
    \boldsymbol{\hat{\rho}}_s
    &= \sum_{t=0}^{n} { n \choose t} (1-\hat p_s )^{n - t} (\hat p_s)^t
    \Bigg[ 0.25\frac{t}{ n}[1 \ 1 \ 1 \ 1]
    + \frac{ n - t}{ n} \boldsymbol{\rho}_s \Bigg] \nonumber \\
    &= \frac{1}{4} \hat p_s [1 \ 1 \ 1 \ 1] + (1-\hat p_s)\boldsymbol{\rho}_s \nonumber \\
    &= \frac{p_s}{3} [1 \ 1 \ 1 \ 1] + \left(1- \frac{4p_s}{3} \right)\boldsymbol{\rho}_s,
\end{align}
which is identical to the analytical constellation update rule for the substitution-only error case given in \eqref{upd_subs2}. Hence, when there are no insertions or deletions, the update rule in \eqref{ID_upd2} yields the exact result for the substitution channel.
\end{remark}

\section{Learning-Enhanced Soft Information Estimation}
\label{sec:ml_llr}

In this section, we introduce learning-based estimators for bit-wise soft information of the codewords corresponding to composite DNA strands. The proposed estimators operate directly on the sequence of observed nucleotide count vectors and aim to improve the robustness of soft information extraction under complex channel impairments.

\subsection{Motivation}

As introduced in the previous section, when IDS errors are present in addition to sampling randomness, synchronization errors lead to variable-length and misaligned channel outputs, rendering the analytical calculation of soft channel outputs intractable. In this case, analytical LLR estimation becomes infeasible for composite DNA data storage systems, since the effective constellation points depend not only on the original design but also on neighboring (and even distant) nucleotides due to the misalignment and dependencies introduced by insertion and deletion errors. 
Furthermore, the IDS channel model considered in the analytical LLR computation represents a simplified abstraction that may not fully capture the biological dependencies and other complex effects observed in practical DNA data storage systems. 
Consequently, only suboptimal analytical constellation update rules can be derived, resulting in approximate LLR calculations, as described previously. Hence, especially in regimes with moderate to high insertion and deletion probabilities, one of the most natural ways to improve robustness is to adopt data-driven solutions for soft information estimation which can also be applied to more complex channel models for which accurate analytical modeling or tractable LLR derivations are not available.

With this motivation, we investigate learning-enhanced soft information estimation strategies. Specifically, we consider two alternative paradigms:

\begin{itemize}
    \item \textit{RhoNet}: A model-based posterior learning approach, where neural networks estimate updated constellation probability vectors, which are subsequently used for analytical LLR computation.

    \item \textit{LLRNet}: A direct LLR learning approach, where bit-wise LLRs are estimated in an end-to-end data-driven manner without explicit constellation updates.
\end{itemize}
While RhoNet preserves the probabilistic structure of the composite DNA errors through analytical post-processing, LLRNet enables a fully end-to-end receiver design by directly producing soft information for channel decoding. In this sense, the proposed learning-enhanced frameworks span the spectrum between classical model-based analytical receivers and fully data-driven inference strategies, offering alternative trade-offs between domain-knowledge integration and end-to-end learning.

\subsection{RhoNet: A Model-Based Posterior Learning Approach} \label{sec:RhoNet}

As a natural extension of our analytical framework, we propose a model-based learning approach in which data-driven techniques are employed solely to update the effective constellation parameters, while LLR computation is performed analytically using the learned constellation parameters with the formulation given in \eqref{eq_llr}.

We consider a composite DNA system with constellation points defined by $\boldsymbol{\rho}_s \in \Delta_L$. Let
\begin{equation}
 \mathbf{\hat{D}} = [ \mathbf{\hat d}_1 \ \mathbf{\hat d}_2 \ \dots \ \mathbf{\hat d}_E] 
 \in \mathbb{R}_{+}^{4 \times E},   
\end{equation} 
where $\mathbf{\hat d}_i \in \mathbb{R}_{+}^4$ for $i \in \{1, 2, \dots, E\}$ represents the observed composite letters after random sampling and the introduction of IDS channel errors. Similar to \eqref{D1}, these composite letters can equivalently be represented as three-dimensional vectors, i.e., $\mathbf{\hat d}_i \in \mathbb{R}_{+}^3$, since their components sum to one; however, we retain the 4-dimensional representation for analytical convenience.

Note that, as will be explained in more detail in the dataset construction section, each read DNA strand is resampled to the original length. This allows us to utilize strands of unequal lengths without introducing additional synchronization degradation. The resulting sequences are used as the network input and are denoted by $\mathbf{D}_{\mathrm{IDS}}$.

Our objective is to estimate the effective channel-dependent constellation parameters associated with the transmitted sequence, which are subsequently incorporated into the analytical framework to compute the bit-wise LLRs. Specifically, let the target soft information vector be represented by
\begin{align}
\boldsymbol{\lambda} = [LLR_{1,1}, \dots, & LLR_{1,L}, \dots, \nonumber \\
&\quad LLR_{E,1}, \dots, LLR_{E,L}] \in \mathbb{R}^N,
\end{align}
where $LLR_{t,b}$ denotes the LLR of the $b$-th bit corresponding to the composite symbol at index $t$, and $N = EL$.

\subsubsection{Bidirectional GRU Architecture}
We employ a BiGRU structure to learn the effective constellation parameters from the composite
letters observed after sampling randomness and IDS channel impairments.
The choice of this architecture is due to the structure of the IDS channel.
Because of insertions and deletions, the channel has memory and the synchronization loss introduces
dependencies among neighboring positions, so that the effective constellation at position $t$
depends on the entire strand rather than on the observation at index $t$ alone. A
feedforward or convolutional estimator cannot represent this
cumulative drift, whereas a recurrent state can, since it is propagated across the
whole sequence.

Furthermore, the soft information required at position $t$ is the posterior of the coded bits
given the entire received strand, and is therefore non-causal. In the classical setting,
such a posterior is computed by a forward-backward recursion over a drift trellis, which
requires explicit modeling of the channel and becomes intractable in the composite DNA setting. A bidirectional recurrent layer has a similar structure with
forward hidden state and the backward
hidden state. The
BiGRU can thus be viewed as a learned, model-free replacement for the forward-backward recursion. Similar bidirectional
recurrent detectors have been adopted for channels with memory in \cite{9279252, 8454325}, and for
insertion/deletion channels in \cite{10622561,  11358926}.

Finally, among gated recurrent cells, we adopt the GRU rather than the long short-term memory
(LSTM) since we have low dimensional count vectors while the strands are of short to moderate length (e.g., $E=160$),
the additional capacity of the LSTM is not required, and the GRU provides the same modeling
ability at a lower computational cost.

As shown in \cite{cho2014learning}, for a forward GRU layer, the hidden state $\overrightarrow{\mathbf{h}}_t$ at time $t$ is updated as
\begin{equation} \label{gru-forward}
    \begin{aligned}
    \mathbf{z}_t &= \sigma\!\left(\mathbf{W}_z [\overrightarrow{\mathbf{h}}_{t-1};\mathbf{D}_{\mathrm{IDS},t}]\right), \\
\mathbf{r}_t &= \sigma\!\left(\mathbf{W}_r [\overrightarrow{\mathbf{h}}_{t-1};\mathbf{D}_{\mathrm{IDS},t}]\right), \\
\tilde{\mathbf{h}}_t &= \tanh\!\left(\mathbf{W}_h [\mathbf{r}_t \odot \overrightarrow{\mathbf{h}}_{t-1};\mathbf{D}_{\mathrm{IDS},t}]\right), \\
\overrightarrow{\mathbf{h}}_t &= (1-\mathbf{z}_t)\odot \overrightarrow{\mathbf{h}}_{t-1} + \mathbf{z}_t \odot \tilde{\mathbf{h}}_t ,
\end{aligned}
\end{equation}
where {$\mathbf{D}_{\mathrm{IDS},t} \in \mathbb{R}_+^4$ denotes the resampled network input $\mathbf{D}_{\mathrm{IDS}}$ at position index $t$, $\sigma(\cdot)$ denotes the sigmoid function and $\odot$ represents element-wise multiplication.

Similarly, the backward GRU processes the sequence in reverse order and produces hidden states $\overleftarrow{\mathbf{h}}_t$ using analogous update equations. A bidirectional GRU layer concatenates the forward and backward hidden states,
\begin{equation}
\mathbf{h}_t = [\overrightarrow{\mathbf{h}}_t \, ; \, \overleftarrow{\mathbf{h}}_t],
\end{equation}
allowing the network to exploit contextual information from both past and future observations.

\subsubsection{Constellation Parameter Update and Analytical LLR Computation}

We follow a supervised learning approach in which channel-dependent effective nucleotide distributions are learned in a data-driven manner. In particular, for each position index $t$ and each candidate constellation point $s\in\{1,\ldots,S\}$, the network estimates an updated probability vector
\begin{equation}
\mathbf{o}_t = \mathbf{W}_{\rho}\mathbf{h}_t + \mathbf{b}_{\rho},
\qquad
\mathbf{o}_t \in \mathbb{R}^{S \times 4},
\end{equation}
followed by a softmax operation applied over the nucleotide dimension:
\begin{equation}
\widehat{\boldsymbol{\rho}}_{t,s}
=
\mathrm{softmax}\!\big(\mathbf{o}_{t,s}\big)
\in \mathbb{R}^{4},
\end{equation}
where 
\begin{equation}
\widehat{\boldsymbol{\rho}}_{t,s}
=
(\hat{\rho}_{t,s,A},\hat{\rho}_{t,s,C},\hat{\rho}_{t,s,T},\hat{\rho}_{t,s,G}),
\end{equation}
resulting in
\begin{equation}
\sum_{b\in\{A,C,T,G\}} \hat{\rho}_{t,s,b}=1.
\end{equation}
Collecting all outputs over a strand of length $E$ yields
\begin{equation} \label{Rhonet-rho}
\widehat{\boldsymbol{\rho}}
=
\left[\widehat{\boldsymbol{\rho}}_{t,s}\right]_{t=1..E,\;s=1..S}
\in \mathbb{R}^{E\times S\times 4}.
\end{equation}

Given the observed composite-letter strands corrupted by IDS errors, these learned constellation parameters are incorporated into the analytical LLR computation framework of \eqref{eq_llr} to obtain the corresponding bit-wise LLRs.

Let $\widehat{\boldsymbol{\lambda}} \in \mathbb{R}^{N}$ denote the resulting LLR vector. To train the network in a supervised manner, we convert these LLRs into logits defined as
\begin{equation}
\widehat{\mathbf{z}} \triangleq -\widehat{\boldsymbol{\lambda}}.
\end{equation}
We then employ the binary cross-entropy (BCE) loss with logits:
\begin{equation}
\mathcal{L}_{\mathrm{BCE}}
=
-\frac{1}{N}
\sum_{j=1}^{N}
\Big[
c_j \log \sigma(\hat{z}_j)
+
(1-c_j)\log\big(1-\sigma(\hat{z}_j)\big)
\Big],
\label{eq:bce_rhonet}
\end{equation}
where $\mathbf{c} \in \{0,1\}^{N}$ denotes the ground-truth coded bit sequence, $N = EL$, $c_j$ is the $j$-th bit of $\mathbf{c}$, $\hat{z}_j$ is the $j$-th element of logit vector $\widehat{\mathbf{z}}$ and $\sigma(\cdot)$ is the sigmoid function.

Hence, the proposed RhoNet architecture does not alter the analytical decoding structure. Instead, it replaces the intractable constellation update rule with learned, index-dependent constellation parameters, which are subsequently used in the standard LLR computation followed by the channel decoder.

\begin{remark}[Hybrid Architecture of RhoNet]
    It is worth noting that RhoNet utilizes the observed sequence $D_{\mathrm{IDS}}$ both as an input to the neural network and within the analytical LLR computation. This structure follows the model-based deep learning paradigm where the BiGRU architecture captures the synchronization loss and position-based dependencies to estimate index-dependent constellations to mitigate the memory effects induced by IDS errors. The analytical block exploits domain knowledge by applying the memoryless multinomial distribution to evaluate the LLR at each position index $t$ using these learned parameters.
\end{remark}

\subsection{LLRNet: A Direct LLR Learning Approach} \label{sec:LLRNet}

We consider the same composite DNA setting as in the previous subsection, but adopt a fully data-driven strategy by estimating the bit-wise soft information directly. In particular, LLRNet eliminates the intermediate constellation-parameter update stage and the subsequent analytical LLR computation in \eqref{eq_llr}. Instead, it produces soft information in an end-to-end manner from the sequence of observed nucleotide count vectors.

Similar to RhoNet, we employ a BiGRU with the same forward and backward operations as in \eqref{gru-forward}. However, in this case, we directly estimate the logits (or, equivalently, the LLRs). Specifically, for each index $t$, a linear projection maps the
final BiGRU output $\mathbf{h}_t$ to a vector of bit-wise logits,
\begin{equation}
\widehat{\mathbf{z}}_t
=
\mathbf{W}_{\mathrm{LLR}} \mathbf{h}_t
+
\mathbf{b}_{\mathrm{LLR}},
\end{equation}
where
$\widehat{\mathbf{z}}_t
=
(\hat z_{t,1},\ldots,\hat z_{t,L})$.
Each logit $\hat z_{t,\ell}$ can be interpreted as an estimate of
\begin{equation}
\hat z_{t,\ell}
\approx
\log
\frac{\sum\limits_{s:u_{s,\ell}=1}
P(\mathbf d_t \mid \boldsymbol{\rho}_s)}
{\sum\limits_{s:u_{s,\ell}=0}
P(\mathbf d_t \mid \boldsymbol{\rho}_s)},
\end{equation}
which corresponds to the log-posterior ratio for bit value $1$
versus $0$.

The corresponding LLRs required by the LDPC decoder are
obtained as
\begin{equation}
\widehat{LLR}_{t,\ell}
\triangleq
-\hat z_{t,\ell}
=
\log
\frac{\sum\limits_{s:u_{s,\ell}=0}
P(\mathbf d_t \mid \boldsymbol{\rho}_s)}
{\sum\limits_{s:u_{s,\ell}=1}
P(\mathbf d_t \mid \boldsymbol{\rho}_s)}.
\end{equation}
Collecting the outputs over all indices yields $\widehat{\mathbf{z}}\in\mathbb{R}^{N}$ with $N=EL$ by vectorizing $(\widehat{\mathbf{z}}_1,\ldots,\widehat{\mathbf{z}}_E)$.

The BiGRU-based estimator is trained in a supervised manner using the ground-truth coded bit sequence $\mathbf{c}\in\{0,1\}^{N}$ as the target with the BCE loss:
\begin{equation}
\mathcal{L}_{\mathrm{BCE}}
=
-\frac{1}{N}
\sum_{j=1}^{N}
\Big[
c_j \log \sigma(\hat{z}_j)
+
(1-c_j)\log\big(1-\sigma(\hat{z}_j)\big)
\Big],
\label{eq:bce_llrnet}
\end{equation}
where $\hat{z}_j$ denotes the $j$-th entry of the vectorized logit output $\widehat{\mathbf{z}}$, and $\sigma(\cdot)$ is the sigmoid function. We note that the mean squared error (MSE) between the estimated and the ground truth LLRs (i.e., the LLRs of the original channel input under pure sampling randomness, without the IDS errors, using the original constellation design) could also be employed as a loss metric. However, our experiments demonstrate that the BCE loss performs better. This is because MSE heavily penalizes numerical deviations from the ground truth even though there is no sign change. In the context of LLRs, such a deviation is largely harmless, and the channel decoder will still perform well. By penalizing these safe magnitude differences, MSE forces the model to optimize for exact numerical matching rather than robust decoding performance. 

Once trained, LLRNet produces soft information that can be directly used by standard channel decoders, without requiring explicit constellation updates or knowledge of channel parameters.

\subsection{Dataset Construction and Training Procedure} \label{sec:dataset_cons}

We construct a synthetic dataset for data-driven LLR estimation.
First, random binary codewords $\mathbf{c}$ are generated and
partitioned into $L$-bit segments, each of which is mapped to a
composite DNA letter according to the predefined composite
constellation set $   \Delta_L = \left\{ \boldsymbol{\rho}_s \in \mathbb{R}_{+}^4 , \; \rho_{s,i} \in [0,1] \,\bigg|\, \sum_{i=1}^{4} \rho_{s,i} = 1 \right\}$.

To model sampling randomness, each composite letter is read
$n$ times, yielding nucleotide count vectors drawn from a
multinomial distribution as described in
Section~\ref{sec:enc_dec}. This process produces the sampled
composite-letter sequence $\mathbf{D}$.

We then introduce channel impairments in the form of
insertions, deletions, and substitutions with probabilities
$p_i$, $p_d$, and $p_s$, respectively, following the IDS
channel model described in Section~\ref{subsec:chs}.
Due to synchronization errors, the resulting strands may have
variable lengths.

To utilize strands of unequal lengths, we apply a length normalization procedure. 
Each received strand is deterministically
re-indexed onto a common grid of $E$ positions via nearest-neighbor index
resampling (i.e., sequence length normalization), so that strands longer than $E$
are decimated at evenly spaced indices and strands shorter than $E$ are expanded by
repeating observed nucleotides. The resulting noisy composite-letter sequence, incorporating
sampling randomness, IDS errors, and length normalization, is
denoted by $\mathbf{D}_{\mathrm{IDS}}$. Under the described system model, the inputs to the training
procedure consist of sequences
$\mathbf{D}_{\mathrm{IDS}}\in\mathbb{R}_{+}^{4\times E}$,
representing resampled nucleotide count vectors corrupted by sampling
randomness and IDS errors.

The desired outputs are the ground-truth coded bit sequences
$\mathbf{c}\in\{0,1\}^{N}$ with $N=EL$.
Training uses the BCE loss as detailed in
Sections~\ref{sec:RhoNet} and~\ref{sec:LLRNet}, where RhoNet
learns constellation updates followed by analytical LLR
computation, while LLRNet directly estimates bit-wise LLRs.

\begin{remark}{(Complexity vs. Performance Trade-off)}
The analytical LLR computation requires marginalizing over the constellation at each position, yielding a complexity of $\mathcal{O}(E \cdot 2^L)$.
In contrast, the learning-based approaches incur a higher base complexity dominated by the recurrent backbone, scaling as $\mathcal{O}(E \cdot H^2)$, where $H$ is the BiGRU hidden dimension.
However, regarding the output layer complexity, LLRNet offers superior scaling. While RhoNet must estimate parameters for every constellation point (scaling as $\mathcal{O}(E \cdot H \cdot 2^L)$), LLRNet decouples the output complexity from the constellation cardinality (scaling as $\mathcal{O}(E \cdot H \cdot L)$).
Ultimately, the increased complexity of the learning-based approaches is justified by the significant performance improvement observed in IDS channels. Unlike the memoryless analytical approach, the BiGRU architecture effectively captures the long-range dependencies and synchronization errors introduced by insertions and deletions significantly outperforming the analytical approximate approach. 
\end{remark}

\section{Numerical Results} \label{sec:numerical_res}

In this section, we evaluate the error rate performance of composite DNA with sampling randomness, and IDS errors. For error correction, we employ three binary LDPC codes with different rates:  

\noindent 1) A rate \(R_1 = 0.3235\) LDPC code based on the 5G standard, obtained from the first base graph with parameters expansion factor \(Z_c = 12\) and set index \(i_{LS} = 1\). The corresponding parity-check matrix \(\mathbf{H}_1\) has size \(552 \times 816\) with $K = 264$ and $N = 816$.  

\noindent 2) A rate \(R_2 = 0.5\) LDPC code from the WRAN (IEEE 802.22) standard, with parity-check matrix \(\mathbf{H}_2\) of size \(240 \times 480\), where $K = 240$ and $N = 480$.  

\noindent 3) A rate \(R_3 = 0.75\) LDPC code from the IEEE 802.11 standard, with parity-check matrix \(\mathbf{H}_3\) of size \(162 \times 648\), where $K = 486$ and $N = 648$.  

\noindent While we focus on the use of LDPC codes, the same framework can be applied with other coding schemes, such as polar codes or convolutional codes.

\begin{table}[t]
\centering
\caption{Constellation points used in simulations.} \label{tab:sim}
\begin{tabular}{|cc|cc|}
\hline
\multicolumn{2}{|c|}{\textbf{$L=3$}}                                                                                                                          & \multicolumn{2}{c|}{\textbf{$L=4$}}                                                                                                                                                                                                       \\ \hline
\multicolumn{1}{|c|}{\begin{tabular}[c]{@{}c@{}}Constellation \\ locations\end{tabular}}   & \begin{tabular}[c]{@{}c@{}}Number of\\ permutations\end{tabular} & \multicolumn{1}{c|}{\begin{tabular}[c]{@{}c@{}}Constellation\\ locations\end{tabular}}                                                                                 & \begin{tabular}[c]{@{}c@{}}Number of\\ permutations\end{tabular} \\ \hline
\multicolumn{1}{|c|}{\begin{tabular}[c]{@{}c@{}}(1,0,0,0)\\ (0.5, 0.5, 0, 0)\end{tabular}} & \begin{tabular}[c]{@{}c@{}}4\\ 4\end{tabular}                    & \multicolumn{1}{c|}{\begin{tabular}[c]{@{}c@{}}(1,0,0,0)\\ (0.5, 0.5, 0, 0)\\ (0.25, 0.25, 0.25, 0.25)\\ (0.333, 0.333, 0.333, 0)\\ (0.666, 0.333, 0, 0)\end{tabular}} & \begin{tabular}[c]{@{}c@{}}4\\ 6\\ 1\\ 4\\ 1\end{tabular}        \\ \hline
\end{tabular}
\end{table}

We consider two composite letter mappings for \(L = 3\) and \(L = 4\), with constellation points specified in Table~\ref{tab:sim}. For \(L = 3\), the points are the same as those in Table~\ref{tab:ex}. It should be emphasized that these points were chosen arbitrarily, without optimization of either the positions or their distribution; such an optimization can be performed as suggested in \cite{kobovich2023m} for improved performance. For LDPC decoding, we apply the log-SPA algorithm with the LLR vector \(\boldsymbol{\lambda}\) with a maximum of 20 iterations. The code repository is available in \cite{Busrategin2026}.

Section \ref{num-analytical} evaluates the analytical LLR computation over the sampling-only, substitution, and ID channels, while Section \ref{num-biGRU} compares RhoNet and LLRNet with the analytical approximation over the IDS channel.

\subsection{Numerical Results with Analytical Approximate LLR Computation} \label{num-analytical}
In this section, we first evaluate the performance of composite DNA under sampling randomness and channel errors using the proposed analytical LLR computations.

Fig. \ref{fig:only_sampling} shows the resulting block error rate (BLER) as a function of the number of DNA samples \(n\). For both \(L = 3\) and \(L = 4\), the BLER decreases as \(n\) increases. With relatively few samples, very low error rates are already achieved, particularly for the rate $R_1 = 0.3235$ code with \(L = 3\). This improvement is due to the larger separation between the constellation points when \(L\) is smaller, which reduces errors caused by sampling randomness. For the code with \(R_2 = 0.5\), the performance slightly degrades, as the higher code rate provides less redundancy. Nevertheless, near-zero BLER can still be achieved with a moderate number of DNA samples. Finally, for \(R_3 = 0.75\), the same trend holds with a further degradation due to the weaker error protection.

Although reducing $L$ (e.g., standard non-composite DNA with $L=2$) increases the minimum distance between the constellation points and hence improves the BLER, composite letters with $L \ge 3$
are essential to exceed the conventional four-base capacity limit and to maximize the logical information density per synthesis cycle. This trade-off mirrors standard digital communication systems. In such systems, higher-order modulation schemes are preferred under favorable channel conditions to maximize data throughput, relying on robust channel codes to mitigate the reduced noise margin.

\begin{figure}[t]
        \centering
%
%
\definecolor{mycolor1}{rgb}{0.63529,0.07843,0.18431}%
\definecolor{mycolor2}{rgb}{0.00000,0.44706,0.74118}%
\definecolor{mycolor3}{rgb}{0.00000,0.49804,0.00000}%
\definecolor{mycolor4}{rgb}{0.87059,0.49020,0.00000}%
\definecolor{mycolor5}{rgb}{0.00000,0.44700,0.74100}%
\definecolor{mycolor6}{rgb}{0.74902,0.00000,0.74902}%
\definecolor{mycolor7}{rgb}{0.47, 0.27, 0.23}

\begin{tikzpicture}

\begin{axis}[%
width=7.35cm,
height=4.2cm,
at={(0.745in,0.444in)},
scale only axis,
unbounded coords=jump,
xmin=0,
xmax=15,
xlabel style={font=\color{white!15!black}},
xlabel={Number of samples (n)},
ymode=log,
ymin=7e-5,
ymax=1,
ylabel style={font=\color{white!15!black}},
ylabel={BLER},
axis background/.style={fill=white},
label style={font=\small},
xmajorgrids,
ymajorgrids,
legend style={at={(0.94,0.6255)},font=\small,nodes={scale=0.75, transform shape},fill opacity=0.2, draw opacity=1,text opacity=1,
legend cell align={left},
        /tikz/column 2/.style={
            column sep=5pt,
        },
    },
xtick={0,2,4,6,8,10,12,14},
xticklabel style={/pgf/number format/fixed, font=\small},
yticklabel style={font=\small},
]

\addplot [color=mycolor1, line width=1.25pt, mark=o, mark options={solid, mycolor1}]
  table[row sep=crcr]{%
1	1\\
2	0.0002\\
3	0\\
4	0\\
5	0\\
};
\addlegendentry{$\text{L = 3, R}_\text{1}\text{ = 0.3235}$}

\addplot [color=mycolor1, dashed, line width=1.25pt, mark=o, mark options={solid, mycolor1}]
  table[row sep=crcr]{%
1	1\\
2	1\\
3	0.916666666666667\\
4	0.0502283105022831\\
5	1.522e-4\\
6	0\\
7	0\\
8	0\\
9	0\\
10	0\\
};
\addlegendentry{$\text{L = 4, R}_\text{1}\text{ = 0.3235}$}

\addplot [color=mycolor2, line width=1.25pt, mark=square, mark options={solid, mycolor2}]
  table[row sep=crcr]{%
1	1\\
2	0.75\\
3	0.00956284153005465\\
4	7.05e-6\\
5	0\\
6	0\\
7	0\\
8	0\\
9	0\\
10	0\\
};
\addlegendentry{$\text{L = 3, R}_\text{2}\text{ = 0.5}$}

\addplot [color=mycolor2, dashed, line width=1.25pt, mark=square, mark options={solid, mycolor2}]
  table[row sep=crcr]{%
1	1\\
2	1\\
3	1\\
4	1\\
5	0.375\\
6	0.0652173913043478\\
7	0.00705171255876427\\
8	6.2789e-04\\
9	8.25e-5\\
10	0\\
};
\addlegendentry{$\text{L = 4, R}_\text{2}\text{ = 0.5}$}

\addplot [color=mycolor4, line width=1.25pt, mark=asterisk, mark options={solid, mycolor4}]
  table[row sep=crcr]{%
1	1\\
2	1\\
3	1\\
4	0.3\\
5	0.0136098509397278\\
6	0.000242670765106255\\
7	0\\
8	0\\
9	0\\
10	0\\
};
\addlegendentry{$\text{L = 3, R}_\text{3}\text{ = 0.75}$}

\addplot [color=mycolor4, dashed, line width=1.25pt, mark=asterisk, mark options={solid, mycolor4}]
  table[row sep=crcr]{%
1	1\\
2	1\\
3	1\\
4	1\\
5	1\\
6	1\\
7	0.954545454545455\\
8	0.65625\\
9	0.338709677419355\\
10	0.170731707317073\\
11	0.061910112359551\\
12	0.021105527638191\\
13	0.00629874025194961\\
14	0.00252191665665906\\
15	0.0007\\
};
\addlegendentry{$\text{L = 4, R}_\text{3}\text{ = 0.75}$}

\end{axis}
\end{tikzpicture}%
        \vspace{-0.75cm}
        \caption{BLER performance for composite DNA data storage system with LDPC code with rate \(R \in \{ 0.3235,0.5,0.75\}\) and $L \in \{3,4\}$.}
        \label{fig:only_sampling}
        \vspace{-0.25cm}
\end{figure}

\begin{figure}[t]
        \centering
        \definecolor{mycolor1}{rgb}{0.63529,0.07843,0.18431}%
\definecolor{mycolor2}{rgb}{0.00000,0.44706,0.74118}%
\definecolor{mycolor3}{rgb}{0.00000,0.49804,0.00000}%
\definecolor{mycolor4}{rgb}{0.87059,0.49020,0.00000}%
\definecolor{mycolor5}{rgb}{0.00000,0.44700,0.74100}%
\definecolor{mycolor6}{rgb}{0.74902,0.00000,0.74902}%
\definecolor{mycolor7}{rgb}{0.47, 0.27, 0.23}
\begin{tikzpicture}
\begin{axis}[%
width=7.35cm,
height=4.2cm,
at={(0.745in,0.444in)},
scale only axis,
unbounded coords=jump,
xmin=0,
xmax=15,
xlabel style={font=\color{white!15!black}},
xlabel={Number of samples (n)},
ymode=log,
ymin=6e-5,
ymax=1,
ylabel style={font=\color{white!15!black}},
ylabel={BLER},
label style={font=\small},
axis background/.style={fill=white},
xmajorgrids,
ymajorgrids,
legend style={align=left,at={(0.96,0.67)},font=\small,nodes={scale=0.75, transform shape},fill opacity=0.52, draw opacity=1,text opacity=1,
legend cell align={left},
        /tikz/column 2/.style={
            column sep=5pt,
        },
    },
xtick={0,2,4,6,8,10,12,14},
xticklabel style={/pgf/number format/fixed, font=\small},
yticklabel style={font=\small},
]
\addplot [color=mycolor1, line width=1.25pt, mark=asterisk, mark options={solid, mycolor1}]
  table[row sep=crcr]{%
1	1\\
2	0.8291\\
3	0.0079\\
4	7.05e-6\\
5	0\\
6	0\\
7	0\\
8	0\\
9	0\\
10	0\\
11	0\\
12	0\\
13	0\\
14	0\\
15	0\\
};
\addlegendentry{\shortstack{Only sampling \\\hspace{-0.35cm}randomness}}
\addplot [color=mycolor2, line width=1.25pt, mark=o, mark options={solid, mycolor2}]
  table[row sep=crcr]{%
1	1\\
2	0.94135\\
3	0.04790\\
4	0.00035\\
5	0\\
6	0\\
7	0\\
8	0\\
9	0\\
10	0\\
11	0\\
12	0\\
13	0\\
14	0\\
15	0\\
};
\addlegendentry{$p_s\text{ = 0.01}$}
\addplot [color=mycolor3, line width=1.25pt, mark=triangle, mark options={solid, rotate=90, mycolor3}]
  table[row sep=crcr]{%
1	1\\
2	0.99990\\
3	0.50205\\
4	0.02480\\
5	0.00055\\
6	0.00003\\
7	0\\
8	0\\
9	0\\
10	0\\
11	0\\
12	0\\
13	0\\
14	0\\
15	0\\
};
\addlegendentry{$p_s\text{ = 0.05}$}
\addplot [color=mycolor4, line width=1.25pt, mark=diamond, mark options={solid, mycolor4}]
  table[row sep=crcr]{%
1	1\\
2	1\\
3	0.95540\\
4	0.31175\\
5	0.02300\\
6	0.00090\\
7	1.2e-5\\
7	0\\
8	0\\
9	0\\
10	0\\
11	0\\
12	0\\
13	0\\
14	0\\
15	0\\
};
\addlegendentry{$p_s\text{ = 0.1}$}
\addplot [color=mycolor7, line width=1.25pt, mark=|, mark options={solid, mycolor7}]
  table[row sep=crcr]{%
1	1\\
2	1\\
3	1\\
4	0.99355\\
5	0.72125\\
6	0.20390\\
7	0.03110\\
8	0.00335\\
9	0.00015\\
10	0\\
};
\addlegendentry{$p_s\text{ = 0.2}$}
\addplot [color=mycolor6, line width=1.25pt, mark=Mercedes star, mark options={solid, mycolor6}]
  table[row sep=crcr]{%
1	1\\
2	1\\
3	1\\
4	1\\
5	1\\
6	1\\
7	1\\
8	0.99840\\
9	0.97365\\
10	0.83500\\
11	0.56725\\
12	0.31195\\
13	0.14020\\
14	0.05655\\
15	0.02015\\
};
\addlegendentry{$p_s\text{ = 0.4}$}

\end{axis}
\end{tikzpicture}%
        \vspace{-0.75cm}
\caption{BLER performance for composite DNA data storage system with $L = 3$ and LDPC code with rate \(R_2 = 0.5\) and substitution channel with $p_s \in \{0.01, 0.05, 0.1, 0.2, 0.4\}$.}
        \label{fig:subs_new}
        \vspace{-0.25cm}
\end{figure}

In Fig.~\ref{fig:subs_new}, we evaluate the performance of the system over a substitution channel using the LDPC code with rate $R_2 = 0.5$, where each sampled strand undergoes random substitutions with probabilities $p_s \in \{0.01, 0.05, 0.1, 0.2, 0.4\}$. The updated constellation points are computed using \eqref{upd_subs}. Comparing the resulting performance with that of the sampling-only channel for $L=3$, we observe that when the substitution probability is small (e.g., $p_s = 0.01$), performance is nearly identical. As $p_s$ increases, the BLER degrades, but for moderate substitution levels, practical LDPC codes still achieve reliable performance. 

To provide a theoretical reference for the curves in Fig. \ref{fig:subs_new}, we also
compute achievable information rates for the substitution channel. Since substitutions cause no
synchronization loss, the output alphabet is finite and can be obtained directly as the set of
count vectors with $\sum_j k_j = n$, and the transition probabilities of the resulting discrete memoryless channel then follow directly from the updated constellation points. We
evaluate the mutual information of this channel under equiprobable inputs, $I_{\mathrm{unif}}(n,p_s)$,
which is the rate at which the LDPC-coded system operates, as well as its capacity $C(n,p_s)$,
obtained with the Blahut-Arimoto algorithm. 
In Fig.~\ref{fig:subs_new}, we have $R_2 L = 1.5$ bits per composite letter. At
$p_s = 0.4$, this rate is achievable with equiprobable inputs only for $n \ge 9$, and the
simulated BLER indeed remains close to one up to $n = 8$ and starts to decay only beyond this
point. The poor performance in this regime is therefore a property of the channel and of the
fixed equiprobable constellation input, rather than a deficiency of the code or of the LLR
computation. At the other extreme, for $p_s = 0.01$ the rate of $1.5$ bits per composite letter
is already achievable with $n^{\star} = 2$ reads, whereas the simulated system requires $n = 4$
to reach a BLER of $10^{-3}$. For $p_s=0.2$, $n=8$ achieves a BLER of $0.00335$ with an effective rate of 
$R_{2}L = 1.5$, while the achievable rate is $\sim2.3$. Hence, we 
operate at $65\%$ of the capacity. This
difference is due to the finite blocklength of the unoptimized LDPC code and the constellation,
and further motivates the design of composite-DNA specific codes and constellations.

\begin{figure}[t]
        \centering
%
%
\definecolor{mycolor1}{rgb}{0.63529,0.07843,0.18431}%
\definecolor{mycolor2}{rgb}{0.10000,0.44706,0.704118}%
\definecolor{mycolor3}{rgb}{0.00000,0.49804,0.00000}%
\definecolor{mycolor4}{rgb}{0.87059,0.49020,0.00000}%
\definecolor{mycolor5}{rgb}{0.00000,0.44700,0.74100}%
\definecolor{mycolor6}{rgb}{0.74902,0.00000,0.74902}%
\definecolor{mycolor2x}{rgb}{0.47, 0.27, 0.23}
\definecolor{mycolor7}{rgb}{0.47, 0.27, 0.23}
\begin{tikzpicture}

\begin{axis}[%
width=7.35cm,
height=4.2cm,
at={(0.745in,0.444in)},
scale only axis,
unbounded coords=jump,
xmin=0,
xmax=15,
xlabel style={font=\color{white!15!black}},
xlabel={Number of samples (n)},
ymode=log,
ymin=6e-5,
ymax=1,
ylabel style={font=\color{white!15!black}},
ylabel={BLER},
label style={font=\small},
axis background/.style={fill=white},
xmajorgrids,
ymajorgrids,
legend style={at={(0.62,0.42)},font=\small,nodes={scale=0.75, transform shape},fill opacity=0.52, draw opacity=1,text opacity=1,
legend cell align={left},
        /tikz/column 2/.style={
            column sep=5pt,
        },
    },
xtick={0,2,4,6,8,10,12,14},
xticklabel style={/pgf/number format/fixed, font=\small},
yticklabel style={font=\small},
]

\addplot [color=mycolor1, line width=1.25pt, mark=square, mark options={solid, mycolor1}]
  table[row sep=crcr]{%
1	1\\
2	0.75\\
3	0.00956284153005465\\
4	7.05e-6\\
5	0\\
6	0\\
7	0\\
8	0\\
9	0\\
10	0\\
};
\addlegendentry{\shortstack{Only sampling \\\hspace{-0.35cm}randomness}}

\addplot [color=mycolor2, line width=1.25pt, mark=o, mark options={solid, mycolor2}]
  table[row sep=crcr]{%
1	1	\\
2	0.92955	\\
3	0.5414	\\
4	0.25545	\\
5	0.1142	\\
6	0.0506	\\
7	0.02425	\\
8	0.0109	\\
9	0.0051	\\
10	0.0025	\\
11	0.0014	\\
12	0.0005	\\
13	0.0003	\\
14	0.00014	\\
15	0	\\
};
\addlegendentry{$\text{p}_\text{i}\text{ = p}_\text{d}\text{ = 0.001}$}

\addplot [color=mycolor4, line width=1.25pt, mark=triangle, mark options={solid, rotate=90, mycolor4}]
  table[row sep=crcr]{%
1	1	\\
2	0.9827	\\
3	0.80825	\\
4	0.5894	\\
5	0.4008	\\
6	0.26275	\\
7	0.16985	\\
8	0.10965	\\
9	0.07135	\\
10	0.04645	\\
11	0.03105	\\
12	0.0201	\\
13	0.01415	\\
14	0.01	\\
15	0.00565	\\
};
\addlegendentry{$\text{p}_\text{i}\text{ = p}_\text{d}\text{ = 0.002}$}

\end{axis}
\end{tikzpicture}%
        \vspace{-0.75cm}
\caption{BLER performance for composite DNA data storage system with $L = 3$ and LDPC code with rate \(R_2 = 0.5\), ID channel with $p_i=p_d =0.001$ and $p_i=p_d =0.002$.} 
        \label{fig:ID_new}
        \vspace{-0.25cm}
\end{figure}

In Fig.~\ref{fig:ID_new}, we consider the ID channel for the same LDPC code with rate $R_2 = 0.5$, where insertions and deletions occur independently with probabilities $p_i = p_d = 0.001$ and $p_i = p_d = 0.002$. For simplicity, only strands of the original length $E$ and $E\pm1$ are retained; however, we note that in all of our numerical results, $n$ represents the total number of sequencing reads, thereby including the discarded strands as well. The LLRs of the composite symbols are calculated using the updated constellation points in \eqref{ID_upd2}. As expected, the BLER is higher than in the sampling-only channel, since insertions and deletions both introduce errors and reduce the effective number of usable strands. Moreover, compared to the substitution error channel, ID errors are more disruptive because insertions and deletions cause index shifts. A single insertion or deletion event affects not only the modified position but also all subsequent bases. This leads to correlated errors across positions, unlike substitution errors which remain independent across nucleotides. Reasonable error rates can still be obtained with a sufficient number of reads, but the discarded-strand fraction grows from approximately 2\% at $p_i = p_d = 0.001$ to 6.3\% at $p_i = p_d = 0.002$, making sequence length variation a primary performance-limiting factor for the analytical approach. This motivates the learning-based receivers of Section \ref{sec:ml_llr}, which retain all received strands via resampling and are evaluated next.

We now compare our approach with the results presented in \cite{9838471}, which considers limited-magnitude probability errors for composite DNA systems and provides both outer and inner bounds, along with practical error-correcting code constructions.  
The underlying channel model is based on limited-magnitude errors and can correct at most a fixed number of $t$ symbol errors along a strand of composite letters.  
On the other hand, under the sampling randomness considered in this work, the majority of composite letters may deviate from the designed constellation points. Therefore, under the error definition in \cite{9838471}, our channel model would classify most positions along the strand as composite letter errors.
As a result, the coding scheme proposed in \cite{9838471}, while effective for limited-magnitude probability errors, performs poorly in our setting due to the fundamental difference in the channel model.

\subsection{Numerical Results with BiGRU-Based Soft Information Estimation} \label{num-biGRU}

For BiGRU-based soft-information estimation over the IDS channel using RhoNet and LLRNet, we consider the same LDPC code of rate $R_2 = 0.5$ introduced previously, with binary codeword length $N = 480$. For all training and testing in this section, we use the constellation points defined earlier for $L = 3$, as summarized in Table~\ref{tab:sim}, which yields $E = N/L = 160$.

We consider two main training procedures with the corresponding dataset construction:
\begin{itemize}
    \item \textbf{Channel State Information (CSI)-robust training:} 
    For a fixed number of samples $n$, we assume that the exact insertion, deletion, and substitution probabilities are not available during training. Accordingly, for the dataset, we generate $B=100{,}000$ codewords with corresponding channel outputs, each with $n$ samples. The insertion, deletion, and substitution probabilities $p_i$, $p_d$, and $p_s$, respectively, are independently and uniformly sampled from the interval $[p_{\text{min}},\, p_{\text{max}}]$ for each sampled DNA strand.
    
    \item \textbf{$n$-robust training:} 
    For predefined and fixed insertion, deletion, and substitution probabilities $p_i$, $p_d$, and $p_s$, respectively, we assume that the exact number of samples per composite letter is not available. Accordingly, for the dataset, we generate $B_n = 20{,}000$ codewords with corresponding channel outputs for each sample number $n \in \{n_{\text{min}}, \ldots, n_{\text{max}}\}$, resulting in a total dataset size of $B = B_n (n_{\text{max}} - n_{\text{min}} + 1)$.
%
\end{itemize}
Both RhoNet and LLRNet are trained using the Adam optimizer with a learning rate of $10^{-3}$ under the BCE loss, with the corresponding BiGRU architectures given in Table~\ref{tab:bigru_llr_160_arch}. Training is performed for up to $50$ epochs with a batch size of $64$, and the model achieving the best validation loss is used for testing. Resampling of the strands is applied during both the training and testing processes.

\begin{table}[t]
\centering
\caption{BiGRU models for RhoNet and LLRNet for window length $E=160$ and constellation size $|\Delta_L|=8$ with $L=3$.}
\label{tab:bigru_llr_160_arch}
\begin{tabular}{l l l}
\hline
\textbf{RhoNet and LLRNet} & \textbf{Layer details} & \textbf{Output shape} \\
\hline
Input & 4 count features per time step & $(160,4)$ \\

Pre-processing & Count normalization layer & $(160,4)$ \\

BiGRU-1 & BiGRU, 256 units, dropout $0.1$ & $(160,512)$ \\

BiGRU-2 & BiGRU, 256 units, dropout $0.1$ & $(160,512)$ \\
\hline
\textbf{RhoNet} &   &   \\
Time-distributed head & &  \\
Reshape & Dense ($8 \times 4$) & $(160,8,4)$ \\
Softmax & Probability normalization & $(160,8,4)$ \\
\hline
\textbf{LLRNet} &   &   \\
Time-distributed head & Dense ($L=3$ logits) & $(160,3)$ \\

\hline
\end{tabular}
\end{table}

\begin{figure}[t]
        \centering
%
%
\definecolor{mycolor1}{rgb}{0.63529,0.07843,0.18431}%
\definecolor{mycolor2}{rgb}{0.10000,0.44706,0.704118}%
\definecolor{mycolor3}{rgb}{0.00000,0.49804,0.00000}%
\definecolor{mycolor4}{rgb}{0.87059,0.49020,0.00000}%
\definecolor{mycolor5}{rgb}{0.00000,0.44700,0.74100}%
\definecolor{mycolor6}{rgb}{0.74902,0.00000,0.74902}%
\definecolor{mycolor2x}{rgb}{0.47, 0.27, 0.23}
\definecolor{mycolor7}{rgb}{0.47, 0.27, 0.23}

\definecolor{c1}{HTML}{0072B2}
\definecolor{c2}{HTML}{E69F00}
\definecolor{c3}{HTML}{009E73}
\definecolor{c4}{HTML}{D55E00}
\definecolor{c5}{HTML}{CC79A7}

\definecolor{cc1}{RGB}{24,73,135}

\definecolor{cc2}{RGB}{93,165,218}

\definecolor{cc3}{RGB}{255,59,48}

\definecolor{cc4}{RGB}{128,32,32}

\definecolor{cc5}{RGB}{31,119,180}

\pgfplotscreateplotcyclelist{my5}{
{cbBlue,      mark=*,         mark options={solid}, mark size=2.2pt},
{cbOrange,    mark=square*,   mark options={solid}, mark size=2.2pt},
{cbGreen,     mark=triangle*, mark options={solid}, mark size=2.4pt},
{cbVermillion,mark=diamond*,  mark options={solid}, mark size=2.2pt},
{cbPurple,    mark=pentagon*, mark options={solid}, mark size=2.4pt},
}

\begin{tikzpicture}

\begin{axis}[%
width=7.35cm,
height=4.2cm,
at={(0.745in,0.444in)},
scale only axis,
unbounded coords=jump,
cycle list name=my5,
xmin=0.001,
xmax=0.01,
xlabel style={font=\color{white!15!black}},
xlabel={$p$},
ymode=log,
ymin=0.1e-6,
ymax=0.3,
ylabel style={font=\color{white!15!black}},
ylabel={BLER},
label style={font=\small},
axis background/.style={fill=white},
xmajorgrids,
ymajorgrids,
yminorgrids,
legend style={
    at={(0.995,0.32)},
    font=\scriptsize,
    nodes={scale=0.95, transform shape},
    fill opacity=0.2,
    draw opacity=1,
    text opacity=1,
    legend cell align={left},
    legend columns=2,   
},
xtick={0.01,  0.009,  0.008,  0.007,  0.006,  0.005, 0.004,  0.003,0.002, 0.001},
xticklabel style={/pgf/number format/fixed, font=\small},
yticklabel style={font=\small},
]

\addplot [line width=1.2pt,color=mycolor1, mark=*, mark options={solid, mycolor1}]
  table[row sep=crcr]{%
0.01	0.967	\\
0.0095	0.955	\\
0.009	0.927	\\
0.0085	0.914	\\
0.008	0.882	\\
0.0075	0.853	\\
0.007	0.77	\\
0.0065	0.743	\\
0.006	0.636	\\
0.0055	0.546	\\
0.005	0.442	\\
0.0045	0.358	\\
0.004	0.2424	\\
0.0035	0.1424	\\
0.003	0.0714	\\
0.0025	0.03494	\\
0.002	0.01006	\\
0.0015	0.00187	\\
0.001	0.00014	\\
};
\addlegendentry{Analytical approximate LLR}

\addlegendimage{empty legend}
\addlegendentry{}

\addplot [color=mycolor2, line width=1.2pt,  mark=triangle*, mark options={solid, mycolor2}]
  table[row sep=crcr]{%
0.01	0.0789	\\
0.0095	0.0558	\\
0.009	0.0395	\\
0.0085	0.028	\\
0.008	0.0176	\\
0.0075	0.0137	\\
0.007	0.006	\\
0.0065	0.0028	\\
0.006	0.0028	\\
0.0055	0.00114	\\
0.005	0.00066	\\
0.0045	0.00028	\\
0.004	0.00015	\\
0.0035	0.000084	\\
0.003	0.000028	\\
0.0025	0.000011	\\
0.002	0	\\
0.0015	0	\\
};
\addlegendentry{RhoNet - CSI-rob: $[0.001,0.01]$}

\addplot [color=mycolor2, line width=1.2pt,  mark=triangle*, dashed, mark options={solid, mycolor2}]
  table[row sep=crcr]{%
0.01	0.1894	\\
0.0095	0.1504	\\
0.009	0.1152	\\
0.0085	0.0863	\\
0.008	0.0606	\\
0.0075	0.0422	\\
0.007	0.0272	\\
0.0065	0.0157	\\
0.006	0.011	\\
0.0055	0.00598	\\
0.005	0.0032	\\
0.0045	0.00148	\\
0.004	0.000768	\\
0.0035	0.000296	\\
0.003	0.000152	\\
0.0025	0.00004	\\
0.002	0.000009	\\
0.0015	0.000002 \\
};
\addlegendentry{$[0.001,0.005]$}

\addplot [color=mycolor4, line width=1.2pt, mark=star, mark options={solid, mycolor4}]
  table[row sep=crcr]{%
0.01	0.0742	\\
0.0095	0.0536	\\
0.009	0.0379	\\
0.0085	0.0286	\\
0.008	0.0169	\\
0.0075	0.0106	\\
0.007	0.0067	\\
0.0065	0.0031	\\
0.006	0.0021	\\
0.0055	0.00128	\\
0.005	0.0006	\\
0.0045	0.00036	\\
0.004	0.000158	\\
0.0035	0.00007	\\
0.003	0.000026	\\
0.0025	0.000009	\\
0.002	0.000001	\\
0.0015	0	\\
};
\addlegendentry{LLRnet - CSI-rob: $[0.001,0.01]$}

\addplot [color=mycolor4, line width=1.2pt, mark=star, dashed, mark options={solid, mycolor4}]
  table[row sep=crcr]{%
0.01	0.1909	\\
0.0095	0.1502	\\
0.009	0.1131	\\
0.0085	0.0873	\\
0.008	0.0622	\\
0.0075	0.042	\\
0.007	0.0271	\\
0.0065	0.0156	\\
0.006	0.0101	\\
0.0055	0.00566	\\
0.005	0.00336	\\
0.0045	0.00152	\\
0.004	0.000858	\\
0.0035	0.000344	\\
0.003	0.000162	\\
0.0025	0.000061	\\
0.002	0.000012	\\
0.0015	3e-6	\\
};
\addlegendentry{$[0.001,0.005]$}

\end{axis}

\end{tikzpicture}%
\caption{BLER performance of the composite DNA data storage system with $L = 3$, 
an LDPC code of rate $R_2 = 0.5$, and $n = 14$ over the IDS channel: 
(1) analytical LLR computation, 
(2) RhoNet, and 
(3) LLRNet with CSI-robust training for 
$[p_{\min},\,p_{\max}] = [0.001,\,0.005]$ 
and $[p_{\min},\,p_{\max}] = [0.001,\,0.01]$.}
 \label{fig:IDS_CSIRobust}
\end{figure}

\begin{figure}[t]
        \centering
%
%

\definecolor{mycolor1}{rgb}{0.63529,0.07843,0.18431}%
\definecolor{mycolor2}{rgb}{0.00000,0.44706,0.74118}%
\definecolor{mycolor3}{rgb}{0.00000,0.49804,0.00000}%
\definecolor{mycolor4}{rgb}{0.87059,0.49020,0.00000}%
\definecolor{mycolor5}{rgb}{0.00000,0.44700,0.74100}%
\definecolor{mycolor6}{rgb}{0.74902,0.00000,0.74902}%

\definecolor{c1}{HTML}{0072B2}
\definecolor{c2}{HTML}{E69F00}
\definecolor{c3}{HTML}{009E73}
\definecolor{c4}{HTML}{D55E00}
\definecolor{c5}{HTML}{CC79A7}

\definecolor{cc1}{RGB}{24,73,135}

\definecolor{cc2}{RGB}{93,165,218}

\definecolor{cc3}{RGB}{255,59,48}

\definecolor{cc4}{RGB}{128,32,32}

\definecolor{cc5}{RGB}{31,119,180}

\begin{tikzpicture}

\begin{axis}[%
width=7.35cm,
height=4.2cm,
at={(0.745in,0.444in)},
scale only axis,
unbounded coords=jump,
xmin=0.003,
xmax=0.01,
xlabel style={font=\color{white!15!black}},
xlabel={$p$},
ymode=log,
ymin=1e-6,
ymax=1e-1,
ylabel style={font=\color{white!15!black}},
ylabel={BLER},
label style={font=\small},
axis background/.style={fill=white},
xmajorgrids,
ymajorgrids,
yminorgrids,
legend style={at={(0.995,0.375)},font=\scriptsize,   legend columns=3,nodes={scale=0.9, transform shape},fill opacity=0.52, draw opacity=1,text opacity=1,
legend cell align={left},
        /tikz/column 2/.style={
            column sep=5pt,
        },
    },
xtick={0.01,  0.009,  0.008,  0.007,  0.006,  0.005, 0.004,  0.003},
xticklabel style={/pgf/number format/fixed, font=\small},
yticklabel style={font=\small},
]

\addlegendimage{empty legend}
\addlegendentry{}

\addplot [color=mycolor1, line width=1.2pt, mark=o, mark options={solid, mycolor1}]
  table[row sep=crcr]{%
0.01	0.0789	\\
0.0095	0.0558	\\
0.009	0.0395	\\
0.0085	0.028	\\
0.008	0.0176	\\
0.0075	0.0137	\\
0.007	0.006	\\
0.0065	0.0028	\\
0.006	0.0028	\\
0.0055	0.00114	\\
0.005	0.00066	\\
0.0045	0.00028	\\
0.004	0.00015	\\
0.0035	0.000084	\\
0.003	0.000028	\\
};
\addlegendentry{IDS: RhoNet} 

\addplot [color=mycolor1, line width=1.2pt, dashed, mark=o, mark options={solid, mycolor1}]
  table[row sep=crcr]{%
0.01	0.0742	\\
0.0095	0.0536	\\
0.009	0.0379	\\
0.0085	0.0286	\\
0.008	0.0169	\\
0.0075	0.0106	\\
0.007	0.0067	\\
0.0065	0.0031	\\
0.006	0.0021	\\
0.0055	0.00128	\\
0.005	0.0006	\\
0.0045	0.00036	\\
0.004	0.000158	\\
0.0035	0.00007	\\
0.003	0.000026	\\
};
\addlegendentry{IDS: LLRNet} 

\addlegendimage{empty legend}
\addlegendentry{}

\addplot [color=mycolor2, line width=1.2pt, mark=star,mark options={solid, mycolor2}]
  table[row sep=crcr]{%
0.01	0.0642	\\
0.0095	0.0459	\\
0.009	0.033275	\\
0.0085	0.022675	\\
0.008	0.0145	\\
0.0075	0.010375	\\
0.007	0.006475	\\
0.0065	0.00385	\\
0.006	0.00196	\\
0.0055	0.00117	\\
0.005	0.00067	\\
0.0045	0.00027	\\
0.004	0.00012	\\
0.0035	0.000057	\\
0.003	0.000031	\\
};
\addlegendentry{ID: RhoNet} 

\addplot [color=mycolor2, line width=1.2pt, dashed, mark=star, mark options={solid, mycolor2}]
  table[row sep=crcr]{%
0.01	0.060525	\\
0.0095	0.044875	\\
0.009	0.0328	\\
0.0085	0.0222	\\
0.008	0.014875	\\
0.0075	0.00955	\\
0.007	0.0059	\\
0.0065	0.00335	\\
0.006	0.0022	\\
0.0055	0.00125	\\
0.005	0.00061	\\
0.0045	0.00032	\\
0.004	0.000128	\\
0.0035	0.000068	\\
0.003	0.000024	\\
};
\addlegendentry{ID: LLRNet} 

\addplot [color=mycolor3, line width=1.2pt, mark=triangle,dotted, mark options={solid,mycolor3}]
  table[row sep=crcr]{%
0.0075	0.1052	\\
0.007	0.0738	\\
0.0065	0.0533	\\
0.006	0.0348\\
0.0055	0.0198	\\
0.005	0.0112	\\
0.0045	0.006	\\
0.004	0.0027	\\
0.0035	0.0008	\\
0.003	0.00026	\\
};
\addlegendentry{IS: Approximate LLR}

\addplot [color=mycolor3, line width=1.2pt, mark=triangle, mark options={solid,mycolor3}]
  table[row sep=crcr]{%
0.01	0.002805	\\
0.0095	0.00205	\\
0.009	0.00149	\\
0.0085	0.001265	\\
0.008	0.000855	\\
0.0075	0.00066	\\
0.007	0.000425	\\
0.0065	0.00043	\\
0.006	0.000313	\\
0.0055	0.000191	\\
0.005	0.000133	\\
0.0045	0.000082	\\
0.004	0.000063	\\
0.0035	0.000034	\\
0.003	0.000011	\\
};
\addlegendentry{IS: RhoNet} 

\addplot [color=mycolor3, line width=1.2pt, dashed,mark=triangle, mark options={solid,mycolor3}]
  table[row sep=crcr]{%
0.01	0.0031	\\
0.0095	0.00279	\\
0.009	0.00206	\\
0.0085	0.00188	\\
0.008	0.001495	\\
0.0075	0.001075	\\
0.007	0.00098	\\
0.0065	0.00069	\\
0.006	0.0005	\\
0.0055	0.00037	\\
0.005	0.000286	\\
0.0045	0.000197	\\
0.004	0.000116	\\
0.0035	0.000054	\\
0.003	0.000025	\\
};
\addlegendentry{IS: LLRNet} 

\addplot [color=mycolor4, line width=1.25pt, mark=|,dotted, mark options={solid, mycolor4}]
  table[row sep=crcr]{%
0.0075	0.1098	\\
0.007	0.0765	\\
0.0065	0.0542	\\
0.006	0.033	\\
0.0055	0.0196	\\
0.005	0.0117	\\
0.0045	0.0059	\\
0.004	0.0023	\\
0.0035	0.0006	\\
0.003	0.00026	\\
};
\addlegendentry{DS: Approximate LLR}

\addplot [color=mycolor4, line width=1.2pt, mark=|,  mark options={solid, mycolor4}]
  table[row sep=crcr]{%
0.01	0.00415	\\
0.0095	0.00312	\\
0.009	0.00259	\\
0.0085	0.00189	\\
0.008	0.001535	\\
0.0075	0.001145	\\
0.007	0.001005	\\
0.0065	0.000725	\\
0.006	0.00061	\\
0.0055	0.000403	\\
0.005	0.00031	\\
0.0045	0.000172	\\
0.004	0.000112	\\
0.0035	0.000061	\\
0.003	0.000032	\\
};
\addlegendentry{DS: RhoNet} 

\addplot [color=mycolor4, line width=1.2pt, mark=|, dashed, mark options={solid, mycolor4}]
  table[row sep=crcr]{%
0.01	0.003595	\\
0.0095	0.00309	\\
0.009	0.00208	\\
0.0085	0.001865	\\
0.008	0.001595	\\
0.0075	0.00116	\\
0.007	0.00092	\\
0.0065	0.00067	\\
0.006	0.0005	\\
0.0055	0.000392	\\
0.005	0.000276	\\
0.0045	0.000172	\\
0.004	0.000118	\\
0.0035	0.000052	\\
0.003	0.000026	\\
};
\addlegendentry{DS: LLRNet} 

\end{axis}
\end{tikzpicture}%
\caption{BLER performance for composite DNA data storage system with $L = 3$, LDPC code with rate \(R_2 = 0.5\), $n=14$ for IDS, ID, IS and DS channels with analytical approximate LLR computation, RhoNet and LLRNet under CSI-robust training with $[p_{\text{min}},\, p_{\text{max}}] = [0.001,\, 0.01]$.} 
        \label{fig:CSIRobust_ID_DS_IS}
\end{figure}

In Fig. \ref{fig:IDS_CSIRobust}, we consider CSI-robust training for both RhoNet and LLRNet and compare the results with the analytical approximate LLR computation based on the constellation update rule, for a fixed sample number $n = 14$. Specifically, we perform training over different CSI ranges to model unknown CSI during training, namely $[p_{\min},p_{\max}] = [0.001,0.005]$ and $[p_{\min},p_{\max}] = [0.001,0.01]$. 
During testing, we set $p_d = p_i = \frac{4}{3}p_s=p$. As expected, both RhoNet and LLRNet, trained over both probability ranges, outperform the analytical approximate LLR computation, which utilizes only strands of lengths $E$ and $E\pm1$.
The gain has two sources: the resampled ML-based solution exploits strands of different lengths as well, and the BiGRU structure exploits the dependencies that insertions and deletions induce among neighboring nucleotides due to synchronization loss. Among the CSI-robust models, the one trained over the interval with smaller IDS probabilities, namely $[p_{\min},p_{\max}] = [0.001,0.005]$, exhibits the poorest performance. In contrast, the CSI-robust model trained over the wider error range $[p_{\min},p_{\max}] = [0.001,0.01]$ achieves superior performance, since a broader CSI interval exposes the model to strands with more severe insertion, deletion, and substitution errors and thereby improves its generalization capability. Finally, RhoNet and LLRNet exhibit very similar BLER performance, since both rely on the same BiGRU backbone and therefore extract soft information of comparable quality, despite RhoNet computing the LLRs analytically from learned constellation parameters while LLRNet producing the bit-wise logits directly. Hence, once the recurrent encoder captures the IDS-induced impairments, the choice between model-based and fully data-driven soft-information estimation has limited impact on the error-rate performance.

In Fig.~\ref{fig:CSIRobust_ID_DS_IS}, we focus on the model from the previous results that was trained over the interval $[p_{\min},p_{\max}] = [0.001,0.01]$ for the IDS channel with $n = 14$, and evaluate its performance under different channel models, namely IDS, insertion-deletion (ID), insertion-substitution (IS), and deletion-substitution (DS). For the IDS, ID, IS, and DS channels, we set $(p_d, p_i, p_s) = (p, p, \frac{3}{4}p)$, $(p_d, p_i, p_s) = (p, p, 0)$, $(p_d, p_i, p_s) = (0, p, \frac{3}{4}p)$, and $(p_d, p_i, p_s) = (p, 0, \frac{3}{4}p)$, respectively, during testing. 
We compare the performance of RhoNet and LLRNet for these channel models with the analytical constellation-update approach. As in the previous case, for all channel models, the BiGRU-based approaches outperform the analytical approximate LLR computation, even though they are trained only for the IDS channel. The highest error rates are observed for the IDS channel, which contains all three error types, closely followed by the ID channel, while the IS and DS channels, each containing only a single type of synchronization error, perform noticeably better. The small gap between the IDS and ID curves indicates that insertions and deletions are the dominant source of degradation in this regime, whereas substitutions, which cause no synchronization loss, have a comparatively limited impact.
We note that the figure does not include the analytical performance curves for the IDS and ID channels, since their performance is significantly inferior compared to the BiGRU-based approaches (e.g., the BLER remains around $10^{-1}$ even when it is reduced to approximately $10^{-3}$ for the learning-based solutions). 
From Figs.~\ref{fig:IDS_CSIRobust} and~\ref{fig:CSIRobust_ID_DS_IS}, we conclude that, even without training separate BiGRU models for different IDS probabilities, CSI-robust training enables improved performance with both RhoNet and LLRNet over a wide range of error probabilities using a single trained model.

\begin{figure}[t]
        \centering
%
%

\definecolor{mycolor1}{rgb}{0.63529,0.07843,0.18431}%
\definecolor{mycolor2}{rgb}{0.00000,0.44706,0.74118}%
\definecolor{mycolor3}{rgb}{0.00000,0.49804,0.00000}%
\definecolor{mycolor4}{rgb}{0.87059,0.49020,0.00000}%
\definecolor{mycolor5}{rgb}{0.00000,0.44700,0.74100}%
\definecolor{mycolor6}{rgb}{0.74902,0.00000,0.74902}%

\definecolor{mycolor1b}{rgb}{0.78117, 0.44706, 0.51059}

\definecolor{mycolor1c}{rgb}{0.92706, 0.81569, 0.83686}

\begin{tikzpicture}

\begin{axis}[%
width=7.35cm,
height=4.2cm,
at={(0.745in,0.444in)},
scale only axis,
unbounded coords=jump,
xmin=3.9,
xmax=12.1,
xlabel style={font=\color{white!15!black}},
xlabel={Number of samples (n)},
ymode=log,
ymin=2e-5,
ymax=0.5,
ylabel style={font=\color{white!15!black}},
ylabel={BLER},
label style={font=\small},
axis background/.style={fill=white},
xmajorgrids,
ymajorgrids,
yminorgrids,
legend style={
    at={(1,1)},
    font=\scriptsize,
    nodes={scale=0.879, transform shape},
    fill opacity=0.52,
    draw opacity=1,
    text opacity=1,
    legend cell align={left},
    legend columns=1,   
},
xtick={2,4,6,8,10,12},
xticklabel style={/pgf/number format/fixed, font=\small},
yticklabel style={font=\small},
]

\definecolor{c1}{HTML}{0072B2}
\definecolor{c2}{HTML}{E69F00}
\definecolor{c3}{HTML}{009E73}
\definecolor{c4}{HTML}{D55E00}
\definecolor{c5}{HTML}{CC79A7}

\definecolor{cc1}{RGB}{24,73,135}

\definecolor{cc2}{RGB}{93,165,218}

\definecolor{cc3}{RGB}{255,59,48}

\definecolor{cc4}{RGB}{128,32,32}

\definecolor{cc5}{RGB}{31,119,180}

\addplot [color=mycolor1, line width=1.25pt, mark=star, mark options={solid, mycolor1}]
  table[row sep=crcr]{%
4	0.13652	\\
5	0.04131	\\
6	0.01409	\\
7	0.00512	\\
8	0.00202	\\
9	0.00104	\\
10	0.00039	\\
11	0.000188	\\
12	0.00008	\\
};
\addlegendentry{RhoNet, perfect CSI}

\addplot [color=mycolor1, line width=1.25pt, dashed, mark=star, mark options={solid, mycolor1}]
  table[row sep=crcr]{%
4	0.14369	\\
5	0.0419	\\
6	0.01414	\\
7	0.00521	\\
8	0.00195	\\
9	0.000841	\\
10	0.0004	\\
11	0.000189	\\
12	0.000071	\\
};
\addlegendentry{RhoNet, CSI rob: $[0.001,0.005]$}

\addplot [color=mycolor1, line width=1.25pt, dotted, mark=star, mark options={solid, mycolor1}]
  table[row sep=crcr]{%
4	0.12967	\\
5	0.03277	\\
6	0.00916	\\
7	0.00314	\\
8	0.001	\\
9	0.00046	\\
10	0.00025	\\
11	0.000076	\\
12	0.00003	\\
};
\addlegendentry{RhoNet, CSI rob: $[0.001,0.01]$}

\addplot [color=mycolor2, line width=1.25pt, mark=o, mark options={solid, mycolor2}]
  table[row sep=crcr]{%
4	0.14723	\\
5	0.04401	\\
6	0.015	\\
7	0.00536	\\
8	0.00258	\\
9	0.00082	\\
10	0.00046	\\
11	0.000209	\\
12	0.000095	\\
};
\addlegendentry{LLRNet, perfect CSI}

\addplot [color=mycolor2, dashed, line width=1.25pt, mark=o, mark options={solid, mycolor2}]
  table[row sep=crcr]{%
4	0.14448	\\
5	0.04437	\\
6	0.01427	\\
7	0.00551	\\
8	0.00253	\\
9	0.000934	\\
10	0.000445	\\
11	0.000237	\\
12	0.000107	\\
};
\addlegendentry{LLRNet, CSI rob: $[0.001,0.005]$} 

\addplot [color=mycolor2, dotted, line width=1.25pt, mark=o, mark options={solid, mycolor2}]
  table[row sep=crcr]{%
4	0.13223	\\
5	0.03514	\\
6	0.01049	\\
7	0.0036	\\
8	0.00127	\\
9	0.00064	\\
10	0.00021	\\
11	0.000104	\\
12	0.00004	\\
};
\addlegendentry{LLRNet, CSI rob: $[0.001,0.01]$}

\end{axis}
\end{tikzpicture}%
\caption{BLER performance of the composite DNA data storage system with $L = 3$ and an LDPC code of rate $R_2 = 0.5$, using RhoNet and LLRNet under $n$-robust training with $n \in \{4,\ldots,12\}$. Three training configurations are considered: (1) fixed error probabilities with $p_d = p_i = \frac{4}{3}p_s = 0.003$; (2) joint CSI- and $n$-robust training over $[p_{\min},\,p_{\max}] = [0.001,\,0.005]$; and (3) joint CSI- and $n$-robust training over $[p_{\min},\,p_{\max}] = [0.001,\,0.01]$. All models are tested at $p_d = p_i = \frac{4}{3}p_s = 0.003$.}
        \label{fig:ID_Krobust_003}
\end{figure}

\begin{figure}[h]
        \centering
%
%

\definecolor{mycolor1}{rgb}{0.63529,0.07843,0.18431}%
\definecolor{mycolor2}{rgb}{0.00000,0.44706,0.74118}%
\definecolor{mycolor3}{rgb}{0.00000,0.49804,0.00000}%
\definecolor{mycolor4}{rgb}{0.87059,0.49020,0.00000}%
\definecolor{mycolor5}{rgb}{0.00000,0.44700,0.74100}%
\definecolor{mycolor6}{rgb}{0.74902,0.00000,0.74902}%

\begin{tikzpicture}

\begin{axis}[%
width=7.35cm,
height=4.2cm,
at={(0.745in,0.444in)},
scale only axis,
unbounded coords=jump,
xmin=3.9,
xmax=12.1,
xlabel style={font=\color{white!15!black}},
xlabel={Number of samples (n)},
ymode=log,
ymin=2e-5,
ymax=1,
ylabel style={font=\color{white!15!black}},
ylabel={BLER},
label style={font=\small},
axis background/.style={fill=white},
xmajorgrids,
ymajorgrids,
yminorgrids,
legend style={
    at={(0.605,0.545)},
    font=\scriptsize,
    nodes={scale=0.9, transform shape},
    fill opacity=0.52,
    draw opacity=1,
    text opacity=1,
    legend cell align={left},
    legend columns=1,   
},
xtick={2,4,6,8,10,12},
xticklabel style={/pgf/number format/fixed, font=\small},
yticklabel style={font=\small},
]


\addplot [color=mycolor1, line width=1.25pt, mark=star, mark options={solid, mycolor1}]
  table[row sep=crcr]{%
4	0.37512	\\
5	0.1672	\\
6	0.07492	\\
7	0.03472	\\
8	0.01681	\\
9	0.00881	\\
10	0.00457	\\
11	0.002378	\\
12	0.001184	\\
};
\addlegendentry{RhoNet, perfect CSI}

\addplot [color=mycolor1, line width=1.25pt, dashed, mark=star, mark options={solid, mycolor1}]
  table[row sep=crcr]{%
4	0.41031	\\
5	0.19832	\\
6	0.09926	\\
7	0.05171	\\
8	0.0284	\\
9	0.01534	\\
10	0.008514	\\
11	0.00492	\\
12	0.002853	\\
};
\addlegendentry{RhoNet, CSI robust: $[0.001,0.005]$}

\addplot [color=mycolor1, line width=1.25pt, dotted, mark=star, mark options={solid, mycolor1}]
  table[row sep=crcr]{%
4	0.3765	\\
5	0.16341	\\
6	0.07263	\\
7	0.03267	\\
8	0.01611	\\
9	0.00766	\\
10	0.00408	\\
11	0.002013	\\
12	0.001141	\\
};
\addlegendentry{RhoNet, CSI robust: $[0.001,0.01]$}

\addplot [color=mycolor2, line width=1.25pt, mark=o, mark options={solid, mycolor2}]
  table[row sep=crcr]{%
4	0.37867	\\
5	0.1713	\\
6	0.07966	\\
7	0.03788	\\
8	0.0178	\\
9	0.00909	\\
10	0.00522	\\
11	0.002545	\\
12	0.001391	\\
};
\addlegendentry{LLRNet, perfect CSI} 

\addplot [color=mycolor2, dashed, line width=1.25pt, mark=o, mark options={solid, mycolor2}]
  table[row sep=crcr]{%
4	0.40961	\\
5	0.2065	\\
6	0.10544	\\
7	0.05428	\\
8	0.02985	\\
9	0.01618	\\
10	0.008685	\\
11	0.004909	\\
12	0.002788	\\
};
\addlegendentry{LLRNet, CSI robust: $[0.001,0.005]$} 

\addplot [color=mycolor2, dotted, line width=1.25pt, mark=o, mark options={solid, mycolor2}]
  table[row sep=crcr]{%
4	0.38291	\\
5	0.17017	\\
6	0.07854	\\
7	0.03636	\\
8	0.01705	\\
9	0.00827	\\
10	0.00432	\\
11	0.002257	\\
12	0.001178	\\
};
\addlegendentry{LLRNet, CSI robust: $[0.001,0.01]$}

\end{axis}
\end{tikzpicture}%
\caption{BLER performance of the composite DNA data storage system with $L = 3$ and an LDPC code of rate $R_2 = 0.5$, using RhoNet and LLRNet under $n$-robust training with $n \in \{4,\ldots,12\}$. Three training configurations are considered: (1) fixed error probabilities with $p_d = p_i = \frac{4}{3}p_s = 0.005$; (2) joint CSI- and $n$-robust training over $[p_{\min},\,p_{\max}] = [0.001,\,0.005]$; and (3) joint CSI- and $n$-robust training over $[p_{\min},\,p_{\max}] = [0.001,\,0.01]$. All models are tested at $p_d = p_i = \frac{4}{3}p_s = 0.005$.}
        \label{fig:ID_Krobust_005}
\end{figure}

In Figs.~\ref{fig:ID_Krobust_003} and~\ref{fig:ID_Krobust_005}, we evaluate the performance of composite DNA over the IDS channel for different numbers of samples, with $n \in \{4,\ldots,12\}$, tested at $p_d = p_i = \frac{4}{3}p_s = 0.003$ and $p_d = p_i = \frac{4}{3}p_s = 0.005$, respectively.
For RhoNet and LLRNet, we consider two different training strategies: (1) $n$-robust training, where the model is trained at a fixed error probability, i.e., $p_d = p_i = \frac{4}{3}p_s = 0.003$ and $p_d = p_i = \frac{4}{3}p_s = 0.005$ for Figs.~\ref{fig:ID_Krobust_003} and~\ref{fig:ID_Krobust_005}, respectively; and (2) CSI- and $n$-robust training, where, in addition to robustness with respect to the number of samples, CSI robustness is incorporated by training the model over the intervals $[p_{\min},p_{\max}] = [0.001,0.005]$ and $[p_{\min},p_{\max}] = [0.001,0.01]$. For both error probabilities, RhoNet and LLRNet consistently outperform the approximate analytical LLR estimation. 
For $p_d = p_i = \frac{4}{3}p_s = 0.003$, as shown in Fig.~\ref{fig:ID_Krobust_003}, the CSI- and $n$-robust model trained over $[p_{\min},p_{\max}] = [0.001,0.005]$ achieves performance very close to the known-CSI case, while the model trained over the wider interval $[p_{\min},p_{\max}] = [0.001,0.01]$ outperforms the $n$-robust model trained at the fixed probability $p_d = p_i = \frac{4}{3}p_s = 0.003$, as the broader training interval again includes strands with higher error levels.
For $p_d = p_i = \frac{4}{3}p_s = 0.005$, as shown in Fig.~\ref{fig:ID_Krobust_005}, the performance of the $n$-robust model trained at the fixed error probability is very close to that of the CSI- and $n$-robust model trained over $[p_{\min},p_{\max}] = [0.001,0.01]$, while it outperforms the model trained over the narrower interval $[p_{\min},p_{\max}] = [0.001,0.005]$, since the fixed training probability already corresponds to a moderate error regime.

\begin{figure}[t]
    \begin{subfigure}{0.5\textwidth}
%
%

\definecolor{mycolor1}{rgb}{0.63529,0.07843,0.18431}%
\definecolor{mycolor2}{rgb}{0.00000,0.44706,0.74118}%
\definecolor{mycolor3}{rgb}{0.00000,0.49804,0.00000}%
\definecolor{mycolor4}{rgb}{0.87059,0.49020,0.00000}%
\definecolor{mycolor5}{rgb}{0.00000,0.44700,0.74100}%
\definecolor{mycolor6}{rgb}{0.74902,0.00000,0.74902}%

\definecolor{c1}{HTML}{0072B2}
\definecolor{c2}{HTML}{E69F00}
\definecolor{c3}{HTML}{009E73}
\definecolor{c4}{HTML}{D55E00}
\definecolor{c5}{HTML}{CC79A7}

\definecolor{cc1}{RGB}{24,73,135}

\definecolor{cc2}{RGB}{93,165,218}

\definecolor{cc3}{RGB}{255,59,48}

\definecolor{cc4}{RGB}{128,32,32}

\definecolor{cc5}{RGB}{31,119,180}

\begin{tikzpicture}

\begin{axis}[%
width=7.35cm,
height=4.2cm,
at={(0.745in,0.444in)},
scale only axis,
unbounded coords=jump,
xmin=0.003,
xmax=0.01,
xlabel style={font=\color{white!15!black}},
xlabel={$p$},
ymode=log,
ymin=1e-6,
ymax=1,
ylabel style={font=\color{white!15!black}},
ylabel={BLER},
label style={font=\small},
axis background/.style={fill=white},
xmajorgrids,
ymajorgrids,
yminorgrids,
legend style={at={(0.995,0.45)},font=\scriptsize,   nodes={scale=0.9, transform shape},fill opacity=0.52, draw opacity=1,text opacity=1,
legend cell align={left},
        /tikz/column 2/.style={
            column sep=5pt,
        },
    },
xtick={0.01,  0.009,  0.008,  0.007,  0.006,  0.005, 0.004,  0.003},
xticklabel style={/pgf/number format/fixed, font=\small},
yticklabel style={font=\small},
]

\addplot [line width=1.2pt,color=mycolor4, mark=|, mark options={solid, mycolor4}]
  table[row sep=crcr]{%
0.01	0.967	\\
0.0095	0.955	\\
0.009	0.927	\\
0.0085	0.914	\\
0.008	0.882	\\
0.0075	0.853	\\
0.007	0.77	\\
0.0065	0.743	\\
0.006	0.636	\\
0.0055	0.546	\\
0.005	0.442	\\
0.0045	0.358	\\
0.004	0.2424	\\
0.0035	0.1424	\\
0.003	0.0714	\\
0.0025	0.03494	\\
0.002	0.01006	\\
0.0015	0.00187	\\
0.001	0.00014	\\
};
\addlegendentry{Analytical approximate LLR}

\addplot [line width=1.2pt,color=mycolor6, mark=diamond, mark options={solid, mycolor6}]
  table[row sep=crcr]{%
0.01	0.8649	\\
0.0095	0.8284	\\
0.009	0.7893	\\
0.0085	0.7459	\\
0.008	0.6819	\\
0.0075	0.6207	\\
0.007	0.5513	\\
0.0065	0.4811	\\
0.006	0.3997	\\
0.0055	0.3232	\\
0.005	0.2541	\\
0.0045	0.181	\\
0.004	0.13368	\\
0.0035	0.08022	\\
0.003	0.04752	\\
};
\addlegendentry{Empirical $\boldsymbol \rho$ estimation (with resampling)}

\addplot [color=mycolor1, line width=1.2pt, mark=o, mark options={solid, mycolor1}]
  table[row sep=crcr]{%
0.01	0.0742	\\
0.0095	0.0536	\\
0.009	0.0379	\\
0.0085	0.0286	\\
0.008	0.0169	\\
0.0075	0.0106	\\
0.007	0.0067	\\
0.0065	0.0031	\\
0.006	0.0021	\\
0.0055	0.00128	\\
0.005	0.0006	\\
0.0045	0.00036	\\
0.004	0.000158	\\
0.0035	0.00007	\\
0.003	0.000026	\\
};
\addlegendentry{RhoNet (resampling - proposed)} 

\addplot [color=mycolor2, line width=1.2pt,  mark=triangle, mark options={solid, mycolor2}]
  table[row sep=crcr]{%
0.01	0.6391	\\
0.0095	0.572	\\
0.009	0.5118	\\
0.0085	0.4482	\\
0.008	0.3665	\\
0.0075	0.3024	\\
0.007	0.2564	\\
0.0065	0.1776	\\
0.006	0.1404	\\
0.0055	0.094	\\
0.005	0.055	\\
0.0045	0.0366	\\
0.004	0.01824	\\
0.0035	0.00804	\\
0.003	0.00268	\\
};
\addlegendentry{RhoNet (without resampling)} 

\addplot [color=mycolor3, line width=1.2pt, mark=star,mark options={solid, mycolor3}]
  table[row sep=crcr]{%
0.01	0.7556	\\
0.0095	0.7154	\\
0.009	0.6631	\\
0.0085	0.6114	\\
0.008	0.5484	\\
0.0075	0.4778	\\
0.007	0.4072	\\
0.0065	0.3456	\\
0.006	0.277	\\
0.0055	0.2054	\\
0.005	0.1515	\\
0.0045	0.0955	\\
0.004	0.06192	\\
0.0035	0.03336	\\
0.003	0.01482	\\
};
\addlegendentry{RhoNet (discard long/short strands)} 

\end{axis}

\end{tikzpicture}%
        \vspace{-0.75cm}
        \caption{BLER performance.}
        \label{fig:CSIRobust_IDS-baseline_bler}
        \vspace{0.5cm}
    \end{subfigure}
    \begin{subfigure}{0.5\textwidth}
%
%

\definecolor{mycolor1}{rgb}{0.63529,0.07843,0.18431}%
\definecolor{mycolor2}{rgb}{0.00000,0.44706,0.74118}%
\definecolor{mycolor3}{rgb}{0.00000,0.49804,0.00000}%
\definecolor{mycolor4}{rgb}{0.87059,0.49020,0.00000}%
\definecolor{mycolor5}{rgb}{0.00000,0.44700,0.74100}%
\definecolor{mycolor6}{rgb}{0.74902,0.00000,0.74902}%

\definecolor{c1}{HTML}{0072B2}
\definecolor{c2}{HTML}{E69F00}
\definecolor{c3}{HTML}{009E73}
\definecolor{c4}{HTML}{D55E00}
\definecolor{c5}{HTML}{CC79A7}

\definecolor{cc1}{RGB}{24,73,135}

\definecolor{cc2}{RGB}{93,165,218}

\definecolor{cc3}{RGB}{255,59,48}

\definecolor{cc4}{RGB}{128,32,32}

\definecolor{cc5}{RGB}{31,119,180}

\begin{tikzpicture}

\begin{axis}[%
width=7.35cm,
height=4.2cm,
at={(0.745in,0.444in)},
scale only axis,
unbounded coords=jump,
xmin=0.003,
xmax=0.01,
xlabel style={font=\color{white!15!black}},
xlabel={$p$},
ymin=0,
ymax=85,
ylabel style={font=\color{white!15!black}},
ylabel={Discard ratio (\%)},
label style={font=\small},
axis background/.style={fill=white},
xmajorgrids,
ymajorgrids,
yminorgrids,
legend style={at={(0.995,0.225)},font=\scriptsize,   nodes={scale=0.9, transform shape},fill opacity=0.52, draw opacity=1,text opacity=1,
legend cell align={left},
        /tikz/column 2/.style={
            column sep=5pt,
        },
    },
xtick={0.01,  0.009,  0.008,  0.007,  0.006,  0.005, 0.004,  0.003},
xticklabel style={/pgf/number format/fixed, font=\small},
yticklabel style={font=\small},
]

\addplot [color=mycolor3, line width=1.2pt, mark=star,dotted, mark options={solid, mycolor3}]
table[row sep=crcr]{%
0.01 76.683286\\
0.0095 75.973857\\
0.009 75.302500\\
0.0085 74.395071\\
0.008 73.446000\\
0.0075 72.440500\\
0.007 71.235500\\
0.0065 69.906143\\
0.006 68.455071\\
0.0055 66.677500\\
0.005 64.739786\\
0.0045 62.432500\\
0.004 59.730357\\
0.0035 56.484857\\
0.003 52.439857\\
};
\addlegendentry{Discard ratio (keep length $E$)}

\addplot [line width=1.2pt,color=mycolor4, mark=|,dotted, mark options={solid, mycolor4}]
table[row sep=crcr]{%
0.01 38.058714\\
0.0095 36.744214\\
0.009 35.466071\\
0.0085 33.937857\\
0.008 32.381500\\
0.0075 30.724643\\
0.007 28.970143\\
0.0065 27.129500\\
0.006 25.136857\\
0.0055 23.056357\\
0.005 20.818071\\
0.0045 18.587429\\
0.004 16.209429\\
0.0035 13.741571\\
0.003 11.228000\\
};
\addlegendentry{Discard ratio (keep length $E$ and $E\pm1$)}

\end{axis}

\end{tikzpicture}%
                \vspace{-0.25cm}
        \caption{Discard ratio.}
        \label{fig:CSIRobust_IDS-baseline_discard}
    \end{subfigure}
    \caption{BLER performance and discard ratio for composite DNA data storage system with $L = 3$, LDPC code with rate \(R_2 = 0.5\), $n=14$, $p_i=p_d=4p_s/3 = p$ for IDS channel with RhoNet and baselines (1) analytical LLR estimation, (2) empirical $\boldsymbol \rho$ estimation (after resampling), (3) no resampling where all strands are used (RhoNet), (4) discarding short/long strands (RhoNet) under CSI-robust training with $[p_{\text{min}},\, p_{\text{max}}] = [0.001,\, 0.01]$.}
    \label{fig:CSIRobust_IDS-baseline}
\end{figure}

Finally, in Fig. \ref{fig:CSIRobust_IDS-baseline}, we evaluate two additional aspects of the proposed framework. First, we consider an empirical baseline in which 1,000,000 composite DNA strands are generated for $L = 3$, $R_2 = 0.5$, and $n = 14$, passed through the identical multinomial IDS channel with error probabilities uniformly sampled from $[p_{\text{min}},\, p_{\text{max}}] = [0.001,\, 0.01]$. The average output nucleotide frequencies are computed for each constellation point after resampling, and these effective constellation vectors are supplied directly to the analytical LLR formulas, eliminating any machine-learning dependency. As shown in Fig. \ref{fig:CSIRobust_IDS-baseline_bler}, this baseline offers only a slight improvement over the analytical approximate LLR and remains far from the learning-enhanced receivers: it captures only global average shifts and cannot track the position-dependent correlations induced by insertions and deletions. 
Second, we isolate the benefit of the resampling step by comparing RhoNet against two alternatives: (i) RhoNet without resampling, where variable-length strands are fed directly into the network, and (ii) RhoNet discarding short/long strands, which processes only strands of the nominal length $E$. The proposed framework outperforms both, since it uses the full pool of received strands while limiting synchronization drift. As seen in Fig. \ref{fig:CSIRobust_IDS-baseline_discard}, at $p_i = p_d = 4p_s/3 = 0.01$ the analytical method discards nearly 40\% of the strands (keeping lengths $E$ and $E \pm 1$), while retaining only length-$E$ strands rejects almost 80\%, confirming that resampling achieves maximum information utilization.

\section{Conclusions} \label{sec:conc}

We study composite DNA data storage systems, which increase storage capacity by using mixtures of composite letters. By drawing an analogy to digital modulation, composite letters were modeled as three-dimensional constellation points on the probability simplex, and the corresponding transition probabilities were defined for a multinomial channel model. To address sampling randomness and synchronization errors, i.e., IDS errors, we proposed the use of practical channel coding schemes.
First, we introduced a tractable approximate solution for LLR computation based on a suboptimal constellation update rule. While this analytical framework remains computationally efficient, it ignores a subset of DNA strands to preserve tractability and operates under specific conditions, leaving room for performance improvements. {\color{black}{To address these limitations, we proposed two machine-learning-based soft information estimation approaches. Specifically, RhoNet performs data-driven constellation point updates followed by analytical LLR computation, thereby retaining a model-based inference structure, whereas LLRNet represents the fully data-driven end of the spectrum by directly estimating the LLRs. 
By leveraging a BiGRU architecture, both models capture dependencies induced by synchronization losses, leading to enhanced soft-information reliability.}} 
Through numerical results using practical binary LDPC codes, we demonstrated that even codes not originally designed for composite letters can achieve high performance in composite DNA data storage systems. Furthermore, we showed that our proposed BiGRU-based approaches remain effective even when the number of samples and CSI information are not explicitly available, by employing robust training strategies.

\section*{Acknowledgement}

This work was funded by the European Union through the ERC Advanced Grant 101054904: “TRANCIDS: Transmission over Channels with Insertions and Deletions.” Views and opinions expressed are, however, those of the authors only and do not necessarily reflect those of the European Union or the European Research Council Executive Agency. Neither the European Union nor the granting authority can be held responsible for them.

\bibliographystyle{IEEEtran}
\flushend
\bibliography{bibs-composite}

@article{kobovich2023m,
  title={{M-DAB}: An Input-Distribution Optimization Algorithm for Composite {DNA} Storage by the Multinomial Channel},
  author={Kobovich, Adir and Yaakobi, Eitan and Weinberger, Nir},
  journal={arXiv preprint arXiv:2309.17193},
  year={2023}
}

@article{shlezinger2023model,
  title={Model-based deep learning},
  author={Shlezinger, Nir and Eldar, Yonina C},
  journal={Foundations and Trends in Signal Processing},
  volume={17},
  number={4},
  pages={291--416},
  year={2023},
  publisher={Emerald Publishing Limited}
}

@ARTICLE{8454325,
  author={Farsad, Nariman and Goldsmith, Andrea},
  journal={IEEE Transactions on Signal Processing}, 
  title={Neural Network Detection of Data Sequences in Communication Systems}, 
  year={2018},
  volume={66},
  number={21},
  pages={5663-5678},
  month = Nov
}

@ARTICLE{9279252,
  author={Zheng, Simeng and Liu, Yi and Siegel, Paul H.},
  journal={IEEE Journal on Selected Areas in Communications}, 
  title={{PR-NN}: {RNN}-Based Detection for Coded Partial-Response Channels}, 
  year={2021},
  volume={39},
  number={7},
  pages={1967-1982},
  month = Jul}

@INPROCEEDINGS{9517821,
  author={Srinivasavaradhan, Sundara Rajan and Gopi, Sivakanth and Pfister, Henry D. and Yekhanin, Sergey},
  booktitle={IEEE International Symposium on Information Theory (ISIT)}, 
  title={Trellis {BMA}: Coded Trace Reconstruction on IDS Channels for {DNA} Storage}, 
  year={2021},
  volume={},
  number={},
  pages={2453-2458},
  month = Jul,
  address = {Melbourne, Australia}
 }

@misc{Busrategin2026,
  author       = {Tegin, Busra},
  title        = {{LearningEnhancedCompositeDNA-IDS}: Source code for `{Learning-Enhanced Composite {DNA} Data Storage Under Sampling Randomness and {IDS} Errors}'},
  year         = {2026},
  publisher    = {GitHub},
  journal      = {GitHub repository},
  howpublished = {\url{https://github.com/busrategin/LearningEnhancedCompositeDNA-IDS.git}}
}

@ARTICLE{11358926,
  author={Kargı, E. Uras and Duman, Tolga M.},
  journal={IEEE Transactions on Communications}, 
  title={Symbol-Level Deep Learning-Based Decoders for Concatenated Codes Over Insertion/Deletion Channels}, 
  year={2026},
  volume={74},
  number={},
  pages={4187-4202},
  month = Jan}

@INPROCEEDINGS{tegin2026robust,
  author={Tegin, Busra and Duman, Tolga M},
  booktitle={IEEE International Conference on Communications}, 
  title={Robust Composite {DNA} Storage under Sampling Randomness, Substitution, and Insertion–Deletion Errors}, 
  year={2026},
  volume={},
  number={},
  pages={1-6},
  month = May,
  address = {Glasgow, United Kingdom}}

@inproceedings{cho2014learning,
  title={Learning phrase representations using {RNN} encoder--decoder for statistical machine translation},
 author={Cho, Kyunghyun and Van Merri{\"e}nboer, Bart and Gulcehre, Caglar and Bahdanau, Dzmitry and Bougares, Fethi and Schwenk, Holger and Bengio, Yoshua},
  booktitle={Proceedings of the 2014 Conference on Empirical Methods in Natural Language Processing (EMNLP)},
  pages={1724--1734},
  year={2014},
  month = Sep
}

@article{grass2015robust,
  title={Robust chemical preservation of digital information on {DNA} in silica with error-correcting codes},
  author={Grass, Robert N and Heckel, Reinhard and Puddu, Michela and Paunescu, Daniela and Stark, Wendelin J},
  journal={Angewandte Chemie International Edition},
  volume={54},
  number={8},
  pages={2552--2555},
  year={2015},
month = Feb,
  publisher={Wiley Online Library}
}

@article{heckel2019characterization,
  title={A characterization of the {DNA} data storage channel},
  author={Heckel, Reinhard and Mikutis, Gediminas and Grass, Robert N},
  journal={Scientific reports},
  volume={9},
  number={1},
  pages={9663},
  year={2019},
month = Jul,
  publisher={Nature Publishing Group UK London}
}

@article{sabary2024reconstruction,
  title={Reconstruction algorithms for {DNA}-storage systems},
  author={Sabary, Omer and Yucovich, Alexander and Shapira, Guy and Yaakobi, Eitan},
  journal={Scientific Reports},
  volume={14},
  number={1},
  pages={1951},
  year={2024},
month = Jan,
  publisher={Nature Publishing Group UK London}
}

@techreport{Wright_IDC_2024_Datasphere,
  author       = {Wright, A.},
  title        = {Worldwide {IDC} Global Datasphere Forecast, 2024--2028: {AI} Everywhere, but Upsurge in Data Will Take Time},
  institution  = {International Data Corporation (IDC)},
  type         = {Technical Report},
  number       = {US52076424},
  year         = {2024}
}

@book{ryan2009channel,
  title={{Channel Codes: Classical and Modern}},
  author={Ryan, William and Lin, Shu},
  year={2009},
  address = {Cambridge,
UK},
  publisher={Cambridge University Press}
}

@INPROCEEDINGS{MGC,
  author={Khabbaz, Ramy and Antonini, Marc and Hanna, Serge Kas},
  booktitle={13th International Symposium on Topics in Coding (ISTC)}, 
  title={Marker Guess \& Check Plus {(MGC+)}: An Efficient Short Blocklength Code for Random Edit Errors}, 
  year={2025},
  volume={},
  number={},
  pages={1-5},
  address= {Los Angeles, CA, USA},
  month = Aug,
  }

@ARTICLE{5733455,
  author={Wang, Feng and Fertonani, Dario and Duman, Tolga M.},
  journal={IEEE Transactions on Communications}, 
  title={Symbol-Level Synchronization and {LDPC} Code Design for Insertion/Deletion Channels}, 
  year={2011},
  volume={59},
  number={5},
  pages={1287-1297},
month = May,
  doi={10.1109/TCOMM.2011.030411.100546}}

@ARTICLE{11107232,
  author={Kobovich, Adir and Weinberger, Nir},
  journal={IEEE Journal on Selected Areas in Information Theory}, 
  title={Input Optimization in the Composite {DNA} Storage Channel}, 
  year={2025},
  volume={6},
  number={},
  pages={248-260},
 }

@ARTICLE{910582,
  author={Davey, M.C. and Mackay, D.J.C.},
  journal={IEEE Transactions on Information Theory}, 
  title={Reliable communication over channels with insertions, deletions, and substitutions}, 
  year={2001},
  volume={47},
  number={2},
  pages={687-698},
month = Feb,
  doi={10.1109/18.910582}}

@inproceedings{chen2003concatenated,
  title={Concatenated codes for deletion channels},
  author={Chen, Johnny and Mitzenmacher, Michael and Ng, Chaki and Varnica, Nedeljko},
  booktitle={IEEE International Symposium on Information Theory},
  pages={218--218},
  year={2003},
month = Jun
}

@ARTICLE{10509546,
  author={Ma, Guochen and Jiao, Xiaopeng and Mu, Jianjun and Han, Hui and Yang, Yaming},
  journal={IEEE Transactions on Communications}, 
  title={Deep Learning-Based Detection for Marker Codes Over Insertion and Deletion Channels}, 
  year={2024},
  volume={72},
  number={10},
  pages={5945-5959},
month = Oct,
  doi={10.1109/TCOMM.2024.3394039}}

@INPROCEEDINGS{10622561,
  author={Kargı, E. Uras and Duman, Tolga M.},
  booktitle={IEEE International Conference on Communications (ICC)}, 
  title={A Deep Learning Based Decoder for Concatenated Coding Over Deletion Channels}, 
  year={2024},
  volume={},
  number={},
  pages={2797-2802},
address = {Denver, CO, USA},
month = Jun,
  doi={10.1109/ICC51166.2024.10622561}}

@ARTICLE{9345504,
  author={Honkala, Mikko and Korpi, Dani and Huttunen, Janne M. J.},
  journal={IEEE Transactions on Wireless Communications}, 
  title={{DeepRx}: Fully Convolutional Deep Learning Receiver}, 
  year={2021},
  volume={20},
  number={6},
  pages={3925-3940},
month = Jun,
  doi={10.1109/TWC.2021.3054520}}

@INPROCEEDINGS{9024433,
  author={Shental, Ori and Hoydis, Jakob},
  booktitle={IEEE Globecom Workshops (GC Wkshps)}, 
  title={Machine {LLR}ning: Learning to Softly Demodulate}, 
  year={2019},
  volume={},
  number={},
  pages={1-7},
address = {Waikoloa, HI, USA},
month = Dec,
  doi={10.1109/GCWkshps45667.2019.9024433}}

@article{erlich2017dna,
  title={{DNA} Fountain enables a robust and efficient storage architecture},
  author={Erlich, Yaniv and Zielinski, Dina},
  journal={Science},
  volume={355},
  number={6328},
  pages={950--954},
  year={2017},
month = Mar,
  publisher={American Association for the Advancement of Science}
}

@article{neiman1965molecular,
  title={On the molecular memory systems and the directed mutations},
  author={Neiman, MS},
  journal={Radiotekhnika},
  volume={6},
  number={1},
  pages={8},
  year={1965}
}

@article{church2012next,
  title={Next-generation digital information storage in {DNA}},
  author={Church, George M and Gao, Yuan and Kosuri, Sriram},
  journal={{Science}},
  volume={337},
  number={6102},
  pages={1628--1628},
  year={2012},
  publisher={American Association for the Advancement of Science},
month = Aug,
}

@article{goldman2013towards,
  title={Towards practical, high-capacity, low-maintenance information storage in synthesized {DNA}},
  author={Goldman, Nick and Bertone, Paul and Chen, Siyuan and Dessimoz, Christophe and LeProust, Emily M and Sipos, Botond and Birney, Ewan},
  journal={Nature},
  volume={494},
  number={7435},
  pages={77--80},
  year={2013},
  publisher={Nature Publishing Group UK London},
month = Jan,
}

@article{anavy2019data,
  title={Data storage in {DNA} with fewer synthesis cycles using composite {DNA} letters},
  author={Anavy, Leon and Vaknin, Inbal and Atar, Orna and Amit, Roee and Yakhini, Zohar},
  journal={Nature biotechnology},
  volume={37},
  number={10},
  pages={1229--1236},
  year={2019},
  publisher={Nature Publishing Group US New York},
month = Sep
}

@INPROCEEDINGS{10619202,
  author={Walter, Frederik and Sabary, Omer and Wachter-Zeh, Antonia and Yaakobi, Eitan},
  booktitle={IEEE International Symposium on Information Theory (ISIT)}, 
  title={Coding for Composite {DNA} to Correct Substitutions, Strand Losses, and Deletions}, 
  year={2024},
  volume={},
  number={},
  pages={97-102},
month = Jul,
address = {Athens, Greece}}

@INPROCEEDINGS{9838471,
  author={Zhang, Wenkai and Chen, Zhen and Wang, Zhiying},
  booktitle={IEEE International Conference on Communications}, 
  title={Limited-Magnitude Error Correction for Probability Vectors in {DNA} Storage}, 
  year={2022},
  volume={},
  number={},
  pages={3460-3465},
  doi={10.1109/ICC45855.2022.9838471},
 month = May,
address = {Seoul, Republic of Korea}
}

@INPROCEEDINGS{10619348,
  author={Cohen, Tomer and Yaakobi, Eitan},
  booktitle={IEEE International Symposium on Information Theory (ISIT)}, 
  title={Optimizing the Decoding Probability and Coverage Ratio of Composite {DNA}}, 
  year={2024},
  volume={},
  number={},
  pages={1949-1954},
  doi={10.1109/ISIT57864.2024.10619348},
month = Jul,
address = {Athens, Greece}}

@article{preuss2024efficient,
  title={Efficient {DNA}-based data storage using shortmer combinatorial encoding},
  author={Preuss, Inbal and Rosenberg, Michael and Yakhini, Zohar and Anavy, Leon},
  journal={Scientific reports},
  volume={14},
  number={1},
  pages={7731},
  year={2024},
  publisher={Nature Publishing Group UK London},
month = May
}

@INPROCEEDINGS{10619334,
  author={Sabary, Omer and Preuss, Inbal and Gabrys, Ryan and Yakhini, Zohar and Anavy, Leon and Yaakobi, Eitan},
  booktitle={IEEE International Symposium on Information Theory (ISIT)}, 
  title={Error-Correcting Codes for Combinatorial Composite {DNA}}, 
  year={2024},
  volume={},
  number={},
  pages={109-114},
  doi={10.1109/ISIT57864.2024.10619334},
month = Jul,
address = {Athens, Greece}}


\end{document}